\documentclass[twocolumn,astrosymb]{aastex63}

\usepackage{subcaption}
\usepackage{multirow}
\usepackage{placeins}
\usepackage{gensymb}
\usepackage{float}
\usepackage{color}
\restylefloat{table}

\newcommand{\sbu}{mag arcsec$^{-2}$}

\newcommand{\Rom}{R\textsc{omulus}}

\definecolor{dgreen}{rgb}{0.0, 0.5, 0.0}

\definecolor{purple}{rgb}{0.6,0.5,0.8}

\received{\today}
\revised{Tomorrow}
\accepted{The day after tomorrow}
\submitjournal{ApJ}

\shorttitle{Quantifying the interplay between definition, orientation and shape of UDGs}
\shortauthors{J. D. Van Nest et al.}

\graphicspath{{./}{figures/}}

\begin{document}

\title{What's in a name?  Quantifying the interplay between the Definition, Orientation, and Shape of Ultra-diffuse Galaxies Using the {\sc Romulus} Simulations}

\correspondingauthor{Jordan D. Van Nest}
\email{jdvannest@ou.edu}

\author{Jordan D. Van Nest}
\affiliation{Homer L. Dodge Department of Physics \& Astronomy, University of Oklahoma, 440 W. Brooks St., Norman, OK 73019, USA}

\author{F. Munshi}
\affiliation{Homer L. Dodge Department of Physics \& Astronomy, University of Oklahoma, 440 W. Brooks St., Norman, OK 73019, USA}

\author{A.C. Wright}
\affiliation{Department of Physics \& Astronomy, Johns Hopkins University, 3400 N. Charles Street, Baltimore, MD 21218, USA}
\affiliation{Department of Physics \& Astronomy, Rutgers, The State University of New Jersey, 136 Frelinghuysen Road, Piscataway, NJ 08854, USA}

\author{M. Tremmel}
\affiliation{Department of Astronomy, Yale University, New Haven, CT 06511, USA}

\author{A.M. Brooks}
\affiliation{Department of Physics \& Astronomy, Rutgers, The State University of New Jersey, 136 Frelinghuysen Road, Piscataway, NJ 08854, USA}
 
\author{D. Nagai}
\affiliation{Department of Physics, Yale University, New Haven, CT 06520, USA}
\affiliation{Department of Astronomy, Yale University, New Haven, CT 06511, USA}

\author{T. Quinn}
\affiliation{Astronomy Department, University of Washington, Box 351580, Seattle, WA, 98195-1580}

\begin{abstract}

We explore populations of ultra-diffuse galaxies (UDGs) in isolated, satellite, and cluster environments using the \Rom25 and \Rom C simulations, including how the populations vary with UDG definition and viewing orientation. Using a fiducial definition of UDGs, we find that isolated UDGs have notably larger semi-major ($b/a$) and smaller semi-minor ($c/a$) axis ratios than their non-UDG counterparts, i.e., they are more oblate, or diskier.  This is in line with previous results that adopted the same UDG definition and showed that isolated UDGs form via early, high-spin mergers.  However, the choice of UDG definition can drastically affect what subsets of a dwarf population are classified as UDGs, changing the number of UDGs by up to $\sim45\%$ of the dwarf population.  We also find that a galaxy's classification as a UDG is dependent on its viewing orientation, and this dependence decreases as environmental density increases. Overall, we conclude that some definitions for UDGs used in the literature manage to isolate a specific formation mechanism for isolated dwarfs, while less restrictive definitions erase a link to the formation mechanism. Thus, how we define UDG populations must be considered if we want to understand the formation and evolution of UDGs.

\end{abstract}


\section{Introduction}
\label{sec:intro}

A significant population of very low surface brightness ($\mu_0 > 24$ \sbu) dwarf galaxies with large effective radii ($>1.5$ kpc) were detected in the Coma cluster by \cite{Gband} and classified as ``ultra-diffuse galaxies’’ (UDGs). While we have known of the existence of low surface brightness (LSB) galaxies for some time \citep[see][]{Disney76,Sandage84,Impey88,Dalcanton97,Conselice18}, hundreds of these particularly diffuse galaxies have been discovered both in cluster environments \citep{Koda15,Mihos15,SBavg,Venhola17,Danieli17} and isolated environments \citep{Papastergis2017,Greco18,Rong19}. Though these galaxies generally have standard dwarf luminosities and metallicities \citep{Greco18,Ferremateu18}, their sizes are more comparable to those of L$_*$ galaxies like the Milky Way. However, other studies argue that these galaxies consist of the extreme end of the continuum of LSBs \citep{Tanoglidis21}, and that their sizes are consistent with those of standard dwarf galaxies when considering the expected location of the gas density threshold for star formation \citep{Chamba20,Trujillo20}.

The recent conclusion that UDGs are so ubiquitous has prompted questions about their formation. Do they originate in standard dwarf mass dark matter halos, or are they failed $L_*$ mass galaxies within larger dark matter halos that for some reason failed to build up their stellar populations? Is their formation driven by internal processes, such as bursty star formation and supernova feedback, or external processes, such as tidal interactions and mergers? Or is it possible that multiple types of UDGs exist, driven by different formation mechanisms? The globular cluster content of UDGs has resulted in both constraining UDGs to live in both `failed' $L_*$ dark matter halos \citep{vanDokkum17,Forbes20, Doppel21} and and also in standard dwarf galaxy dark matter halos \citep{Amorisco18,Saifollahi20, Carleton21}. Some measurements even indicate UDGs with largely undermassive dark matter halos \citep{vanDokkum18,Danieli19,vanDokkum19}. 

Idealized simulations \citep[e.g.][]{Yozin15,Chowdhury19}, analytic and semi-analytic models \citep[e.g.][]{Amorisco16,Rong19,Ogiya18,Carleton19}, and cosmological simulations \citep[e.g.][]{Reff,Chan18,Jiang19,Liao19,Martin19} have all been utilized in order to better understand UDG formation, resulting in multiple channels for their formation, including:  dynamical origins, with UDGs forming through tidal heating and/or stripping from interactions \citep[e.g.][]{Ogiya18, Jiang19, Liao19}; 
 bursty star formation and supernova feedback, resulting in extended, diffuse stellar distributions and UDG-like properties \citep[e.g.][]{Reff,Chan18,Martin19}; ram pressure stripping, resulting in a extended, passively evolving stellar populations in cluster galaxies \citep{Tremmel20}; and, finally, mergers, resulting in isolated UDG progenitors with extended effective radii, high spin, and low central star formation \citep{Wright21}.

The above demonstrates that theory predicts multiple formation channels for UDGs, though conclusions may limited by both physical models in simulations and limited statistics in observations. However, there is no consensus about what constitutes the `ultra-diffuse' designation; the questions of `how dim' and `how large' are all answered differently by different groups. Some identify UDGs by the central surface brightness \citep[e.g.][]{Gband,Forbes20}, while others use the effective surface brightness \citep[e.g.][]{Reff,CardonaBarrero20} or the average surface brightness within the effective radius \citep[e.g.][]{Koda15,Leisman17,Martin19}. Many groups require the effective radii of UDGs to be larger than 1.5 kpc \citep[e.g.][]{Gband,Leisman17,Forbes20} while others only require them to be larger than 1 kpc \citep[e.g.][]{Reff,CardonaBarrero20}. The question then becomes: are we comparing apples to apples? Is each group identifying the same population of galaxies as ultra-diffuse? Is this consistent across environment? If not, does it matter?

To further complicate this question, we can consider the orientation of the galaxy when fitting surface brightness profiles. While observers are limited to observing galaxies from Earth, it is standard practice in simulations to orient galaxies to face-on positions before identifying UDGs in order to maximize the sample size \citep{Reff,Jiang19,Liao19,Tremmel20,Wright21}.  But what role does orientation play in identifying a UDG? Does this vary with environment? If a dependence on orientation exists, it would stem from the morphology of the galaxies. Intuitively, one would expect a largely spherical galaxy to appear roughly equivalent at all viewing angles, whereas a `disky' galaxy's appearance would be very dependent on orientation. As with the criteria for being ``ultra-diffuse'', there is no cohesive understanding about the shapes of UDG populations. Some groups speculate that isolated UDGs favor a prolate morphology \cite[e.g.][]{Burkert17,Jiang19}, while others claim an oblate-triaxial morphology is preferred \cite[e.g.][]{Rong20,KadoFong21}. Identifying a cohesive shape distribution for UDGs could provide insights into how they evolve and what formation channels exist.

In this work, we study the populations of galaxies that common definitions of UDGs identify, and whether they are consistent across definitions.  We also explicitly test what effect the orientation of galaxies has on the UDG populations in the \Rom25 and \Rom C simulations.  Using the large sample of dwarfs and UDGs in the Romulus simulations, which contain varying environments from field to cluster, we have the ability to explicitly test the above questions.  In doing so, we demonstrate that the physical processes separating UDGs from the underlying dwarf population can be explicitly dependent on definition.  Thus, it is imperative to consistently identify UDGs in order to solve the mystery of their origin. 

The paper is organized as follows. In Section \ref{sec:simulations}, we detail the \Rom\ simulations and our sample of 1249 resolved dwarf galaxies in isolated, satellite, and cluster environments, with anywhere from 354 to 990 UDGs (depending on how they are identified). In Section \ref{sec:shapes}, we study the shapes of the dwarf galaxies in different environments, and the correlation between a galaxy's shape evolution and its status as a UDG. In Section \ref{sec:effects} we explore the effects of changing the UDG criteria and the galaxies' orientations on the resultant UDG population and our results in Section \ref{sec:shapes}. Finally, we summarize our findings in Section \ref{sec:conclusions}.

\section{The Romulus Simulations}
\label{sec:simulations}

All galaxies analyzed in this work come from the \Rom25 \citep{Rom25} and \Rom C \citep{RomC} simulations. These are high-resolution cosmological simulations run using C\textsc{ha}NG\textsc{a}, an $N$-body+Smooth Particle Hydrodynamics (SPH) code \citep{Changa}. C\textsc{ha}NG\textsc{a} implements several physics models from its predecessor G\textsc{asoline} \citep{Gasoline}; however, its improved SPH implementation allows for better capture of fluid instabilities through reduced artificial surface tension \citep{Gasoline2}. Physics below the resolution limit, such as stellar formation and feedback, as well as supermassive black hole (SMBH) growth and feedback, are governed by sub-grid prescriptions.

The \Rom25 simulation is a 25 Mpc-per-side uniform volume simulation. The simulated galaxies match important $z=0$ scaling relations, including the stellar mass-to-halo mass relation \citep{SM-HM} and the stellar mass-to-SMBH mass relation \citep{SM-SMBH}. Furthermore, \citet{Rom25} demonstrated the simulation's ability to produce realistic galaxies across four orders of magnitude in halo mass, ranging from dwarf galaxies resolved with over 10,000 particles to groups, while reproducing observations of high-redshift SMBH growth and star formation. 

The \Rom C simulation is a cosmological zoom-in simulation of a galaxy cluster. At $z=0$ the cluster has an $R_{200}$ (the radius at which the enclosed density is 200 times the critical density of the universe) of 1033 kpc and an $M_{200}$ (the mass enclosed within $R_{200}$) of $1.15\times10^{14}$ $M_\odot$ . The initial conditions for the cluster were extracted from a 50 Mpc-per-side uniform volume simulation using the renormalization technique of \citet{ZoomReNorm}. 

Both of the \Rom\ simulations are evolved to $z=0$ with a $\Lambda$CDM cosmology ($\Omega_0=0.3086,\Lambda=0.6914,h = 0.6777, and \sigma_8 = 0.8288$) following the \citet{Planck}. Gravitational interactions are resolved with a spline force softening length of 350 pc, which is a Plummer equivalent of 250 pc, that converges to Newtonian force at 700 pc. The simulations oversample dark matter particles, such that the initial high-resolution dark matter particle count is 3.375 times that of the gas particles. This oversampling results in dark matter and gas particles with similar masses, $3.39\times10^5$ $M_\odot$ and  $2.12\times10^5$ $M_\odot$ respectively. The similar particle masses aid in reducing the numerical effects from two-body scattering and energy equipartition, both of which can lead to spurious growth in galaxy sizes within simulations \citep{Ludlow}. The increased resolution in dark matter also gives the simulations the ability to track the dynamics of SMBHs within galaxies \citep{TremmelSMBH}. Additionally, the simulations are allowed more realistic treatment of weak and strong shocks \citep{Gasoline2} through an on-the-fly time-step adjustment and time-dependent artificial viscosity \citep{Saitoh09}. The sub-grid parameters have been optimized to create  galaxies across a halo mass range of $10^{10.5-12}$ $M_\odot$, and an updated implementation of turbulent diffusion \citep{Gasoline2} allows for the formation of a realistic metal distribution within galaxies \citep{Shen10} and the intracluster medium \citep{Wadsley08,Butsky19,RomC}.

\subsection{Sub-grid Physics and Star Formation}
\label{sec:subgrid}
To approximate reionization effects, the \Rom\ simulations include a cosmic UV background \citep{Haardt12} with self-shielding from \citet{Pontzen08}. The simulations implement primordial cooling for neutral and ionized H and He. This cooling is calculated from H and He line cooling \citep{Cen92}, photoionization, radiative recombination \citep{Black81,Verner96}, collisional ionization rates \citep{Abel97}, and bremsstrahlung radiation. Although the \Rom\  simulations are high resolution, they lack the ability to resolve the multiphase interstellar medium (ISM) or track the creation and annihilation of molecular hydrogen. It has been shown that simulating metal-line cooling at low resolution and without the presence of molecular hydrogen physics can lead to overcooling in spiral galaxies \citep{Christensen14}, thus the simulations do not include high-temperature metal-line cooling (see \citet{RomC} for more details of this omission). Low-temperature metal-line cooling is implemented following \citet{Bromm01}.

Star formation (SF) in the \Rom\ simulations is a stochastic process governed by sub-grid models. In simulations of this resolution \citep{Stinson06}, SF is regulated by parameters that determine the physical requirements for gas to become star forming, the efficiency of the SF, and the coupling of supernova (SN) energy to the ISM:
\begin{itemize}
    \item Gas must have a minimum density of n$_{\text{SF}}=0.2\text{ cm}^{-3}$ and a temperature of no greater than T$_{\text{SF}}=10^4$ K in order to form stars.
    \item The probability $p$ of a star particle forming from a gas particle with dynamical time $t_{\text{dyn}}$ is given by:
    \begin{equation}
        p = \frac{m_{\text{gas}}}{m_{\text{star}}}(1-e^{-c_{\text{SF}}\Delta t/t_{\text{dyn}}})
    \end{equation}
    where $c_{\text{SF}}$ is the star-forming efficiency factor (here set to 0.15) and $\Delta t$ is the formation timescale (here set to $10^6$ yr).
    \item The fraction of the canonical 10$^{51}$ erg SN energy coupled to the ISM is $\epsilon_{\text{SN}} = 0.75$.
\end{itemize}
When star particles are formed, they have a mass equivalent to 30\% of the initial gas particle mass, i.e., $M_\star=6\times10^4$ $M_\odot$, and represent a single stellar population with a \citet{Kroupa01} initial mass function.
Stars whose masses are in the range of 8-40 $M_\odot$ undergo Type II SNe that are implemented using `blastwave' feedback, following \citet{Stinson06}. The SN injects thermal energy into nearby gas particles, where the adiabatic expansion phase is replicated by temporarily disabling cooling. Mass and metal diffusion into the ISM follow \citet{Shen10} and \citet{Governato15}. Stars whose masses exceed 40 $M_\odot$ are assumed to collapse directly into a black hole.

The \Rom\ simulations include a novel implementation of black hole physics \citep{Bellovary10,TremmelSMBH,Rom25,Tremmel18a,Tremmel18b,RomC}. To ensure that SMBHs are seeded in gas collapsing faster than the star formation or cooling time scales, they only form in regions of pristine (Z $<3\times10^{-4}$ Z$_\odot$ for \Rom25 and Z $<10^{-4}$ Z$_\odot$ for \Rom C), dense ($n > 15n_{\text{SF}}$), and cool ($9.5\times10^3 <$ T $< 10^4$ K) gas. The seed mass is $10^6$ $M_\odot$ and most SMBHs form within the first Gyr of the simulation. An implementation of a dynamical friction sub-grid model \citep{TremmelSMBH} allows the tracking of SMBH orbits as they move freely within their host galaxies. The growth of SMBHs is modeled via a modified Bondi-Hoyle accretion formalism that accounts for gas supported by angular momentum. Feedback from an active galactic nucleus is approximated by converting a fraction (0.2\%) of the accreted mass into thermal energy and distributing it to the surrounding gas particles.

\subsection{Halo Identification}
\label{sec:haloid}

The \Rom\ simulations use the Amiga Halo Finder \citep[AHF;][]{AHF} to identify dark matter halos, subhalos, and their associated baryonic content. AHF uses a spherical top-hat collapse technique to determine the each halo's virial mass ($M_{vir}$) and radius ($R_{vir}$), following a procedure similar to \citet{BryanNorman98}. Halos are considered resolved if they have a  virial mass of at least $3\times10^9$ $M_\odot$, corresponding to a dark matter particle count of $\sim10^4$, and a stellar mass of at least $10^7$ M$_{\odot}$, corresponding to a star particle count of $\sim150$, at z=0. Galaxies are considered to be resolved dwarf galaxies if they meet the above criteria, as well as have a stellar mass less than 10$^9$ $M_\odot$. When calculating stellar masses, we use photometric colors, following \citet{Munshi13} to better represent the values inferred from typical observational  techniques. In \Rom C, we identify a sample of 201 dwarf galaxies in a cluster environment. In \Rom25, we have a sample of 377 dwarf galaxies in a satellite environment, where a galaxy is classified as a satellite if its center is within the virial radius of a larger halo. While we implement no restrictions on the satellite-to-host mass ratios when determining satellites, a check of our satellite population at z=0 shows that we avoid considering any significant fringe cases, such as two nearly equivalent mass halos merging. Of our 377 satellites, only 19 have a satellite-to-host mass ratio greater than 1:10, and only three have a ratio greater than 1:3.  We also identify a sample of  671 dwarf galaxies in an isolated environment in \Rom25. The  criteria for being isolated follows from \citet{Geha12}, and requires being neither a satellite of another halo nor within 1.5 Mpc of any galaxy with $M_{\star} > 2.5\times10^{10}$ $M_\odot$. Galaxies are analyzed using  P\textsc{ynbody} \citep{pynbody} and tracked across time using T\textsc{angos} \citep{Tangos}, both of which are publicly available.

\subsection{UDG Identification}
\label{subsec:UDG_ID}

The process of identifying UDGs follows the previous studies of UDGs in the \Rom\ simulations \citep{Tremmel20,Wright21}. When analyzing a galaxy, we first orient it in a face-on position based on the angular momentum. This calculation is performed on the gas particles within the inner 5 kpc of the galaxy, or the star particles when less than 100 gas particles are present. Once oriented, we generate azimuthally averaged surface brightness profiles with 300 pc sampling (the spatial resolution of the Romulus simulations). We generate profiles in Johnson $B$, $V$, and $R$ bands, and calculate $g$-band surface brightness from the $B$ and $V$ bands, following \citet{Jester}. We then fit a S\'ersic profile through all points brighter than 32 \sbu, the general depth of sensitive observations \citep{Trujillo16,Borlaff19}. The S\'ersic profile is defined as
\begin{equation}
    \mu(r) = \mu_{\text{eff}}+2.5(0.868n-0.142)\left(\left(\frac{r}{r_{\text{eff}}}\right)^{1/n}-1\right)
\end{equation}
where $\mu(r)$ is the surface brightness at radius r, $r_{\text{eff}}$ is the effective radius (half-light radius), $\mu_{\text{eff}}$ is the effective surface brightness (surface brightness at the effective radius), and $n$ is the S\'ersic index \citep{Sersic}. When fitting a S\'ersic profile, we allow the effective surface brightness to range between 10 and 40 \sbu, the effective radius to range between 0 and 100 kpc, and the S\'ersic index to range between 0.5 and 16.5.  In Section \ref{sec:effects} we explore different definitions of UDGs, but we first adopt our fiducial definition in Section \ref{sec:shapes}.  For our fiducial definition, a halo is considered ultra-diffuse if its central $g$-band surface brightness is dimmer than 24 \sbu\ and its effective radius is larger than 1.5 kpc, following \citet{Gband}.

\begin{figure}
  \centering
  \includegraphics[width=\linewidth]{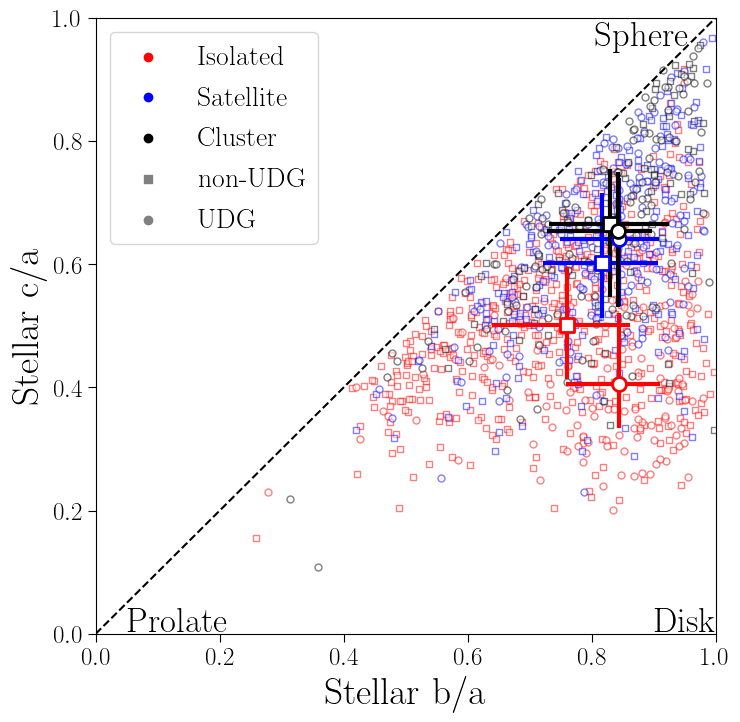}
\caption{The $c/a$ axis ratios plotted against the $b/a$ axis ratios for the dwarf galaxies' stellar distributions at $z=0$ in various environments. The populations are separated into UDGs and non-UDGs according to Section \ref{subsec:UDG_ID}. The bold points represent the medians of the populations, while the error bars show the 25$^{\text{th}}$ to 75$^{\text{th}}$ percentiles. Isolated galaxies, in general, are `diskier' in morphology than the satellite and cluster galaxies, though isolated UDGs are the most `disky' population.}
\label{fig:CvB}
\end{figure}

\begin{figure*}
  \centering
  \includegraphics[width=\linewidth]{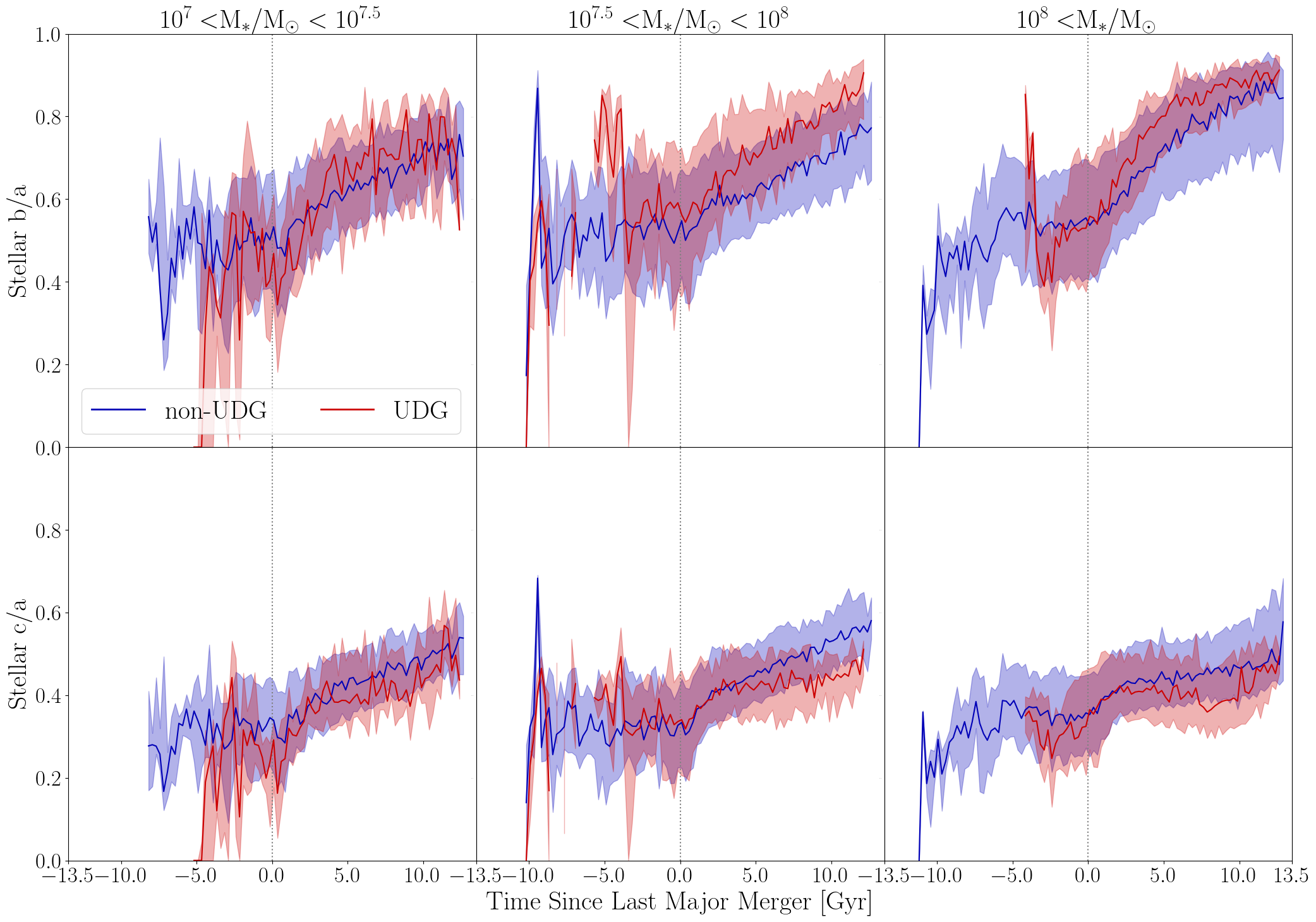}
\caption{The isolated galaxies' stellar $b/a$ and $c/a$ axis ratios as a function of the time elapsed since their final major merger ($t=0$). The shaded regions represent the 25$^{\text{th}}$ to 75$^{\text{th}}$ percentiles of the UDG (red) and non-UDG (blue) populations. Only time bins with at least five data points are plotted. Intermediate- and high-mass UDGs evolve to higher $b/a$ values and lower $c/a$ values than their non-UDG counterparts.}
\label{fig:Field-StellarOnlyShape}
\end{figure*}

\begin{figure*}
\centering
\includegraphics[width=\linewidth]{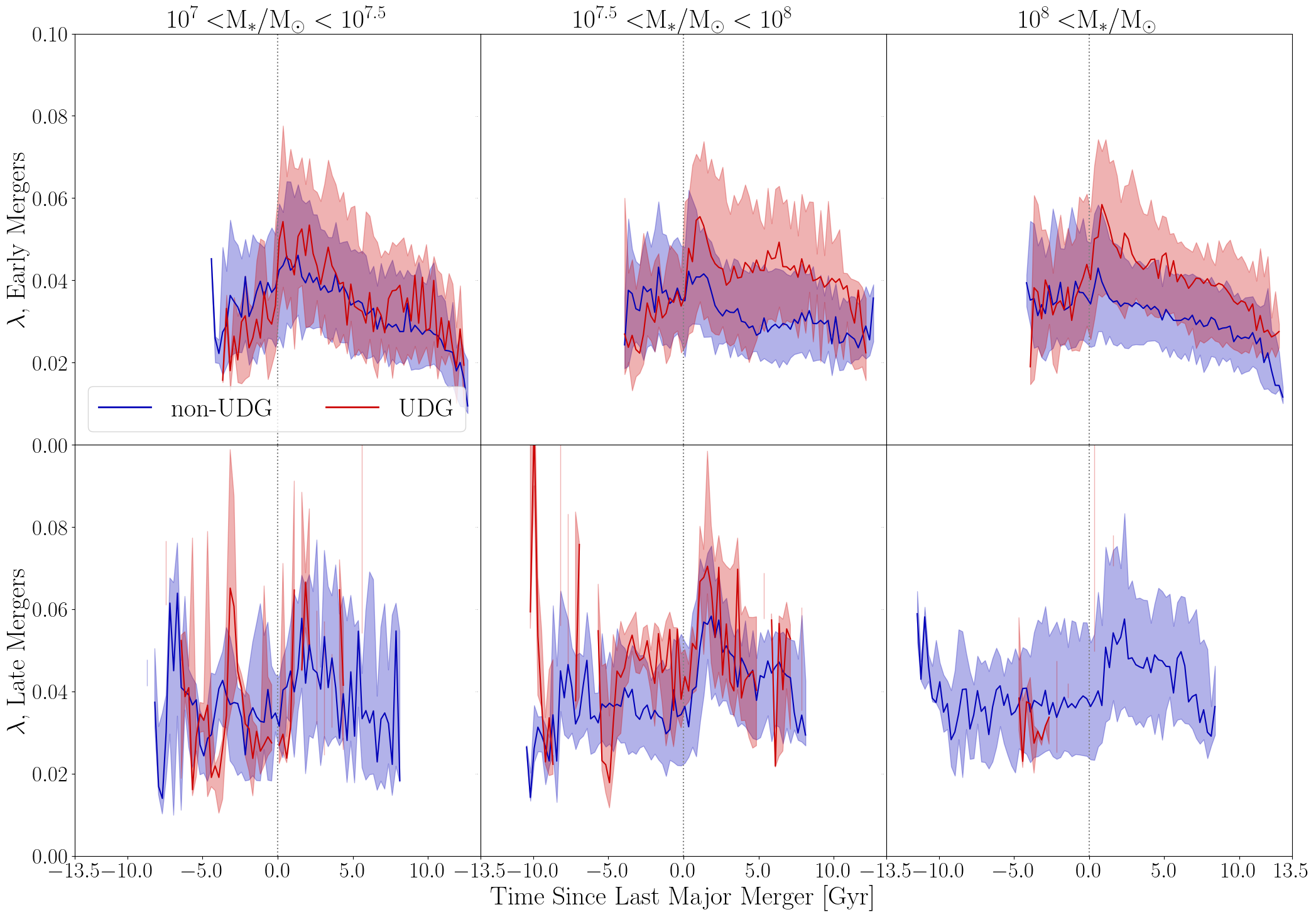}
\caption{The spin parameter, $\lambda$, for isolated galaxies separated into UDGs and non-UDGs for the fiducial definition. The top row contains galaxies that finished merging within the first 5~Gyr of the simulation, while the bottom row contains galaxies with major mergers after 5~Gyr. The shaded regions represent the 25$^{\text{th}}$ to 75$^{\text{th}}$ percentiles. Only time bins with at least five data points are plotted. Intermediate- and high-mass, early-merging UDGs evolve to higher spin values after merging. While intermediate- and high-mass late-merging galaxies evolve to similarly high spins, they do not typically result in UDGs.
}
\label{fig:Lambda}
\end{figure*}

\section{Shapes of UDGs and Non-UDGs}
\label{sec:shapes}

To study the possible differences between UDG and non-UDG morphology in different environments, we analyzed the $b/a$ and $c/a$ axis ratios for all dwarf galaxies, which correspond to the major axis ratios when looking at the galaxies from face-on and edge-on orientations, respectively. Hence, a $b/a\approx c/a\approx1$ would be a spherical distribution, a $b/a\gg c/a$ would be an oblate (or `disky') distribution, and a $b/a\approx c/a<1$ would be a prolate distribution. Axis ratios are derived from the eigenvalues of the shape tensor, which is calculated by dividing the moment of inertia tensor by the total mass. Shape tensors are generated using star particles that are separated into radial bins. The values presented in this work come from the stellar particle bin located at twice the half-light radius. 

Figure \ref{fig:CvB} shows the stellar $c/a$ axis ratios plotted against the stellar $b/a$ axis ratios for dwarf galaxies in all environments at $z=0$. In general, we see that the isolated galaxies reside closer to the lower right-hand corner of the plot, indicating a more disk-like structure. We find that in the cluster and satellite environments, the $b/a$ and $c/a$ axis ratios for UDGs are in good agreement with non-UDGs, but the isolated galaxies show a notable disparity between the two populations. In general, the isolated UDG population is more oblate than the underlying dwarf population. 

Our results seem to contrast with those of \citet{Jiang19}, who found that field UDGs are more prolate than non-UDGs. However, the low-mass non-UDG axis ratios from Figure 4 of \citet{Jiang19} ($b/a\approx.7$, $c/a\approx.414$) are well within the errors of our isolated non-UDGs.  While we are unsure why our UDG results differ, there are some probable causes that can be ruled out. In this work, we perform shape measurements using stellar particles in a bin around two half-light radii, while \citet{Jiang19} used all stellar particles within one half-light radius. However, if we calculate our shapes using a similar method, we still find that our isolated UDGs are more oblate than the non-UDGs. Another possible explanation for our differing results is the \Rom\ simulations' inability to form cores. The \Rom\ simulations do not resolve the multiphase ISM, thus stars must be allowed to form in gas that is relatively diffuse. As a result, the galaxies lack the bursty, clustered central star formation required to produce feedback strong enough to create a core in the dark matter halo \citep[e.g.,][]{Dutton2020}. However, in performing our shape analysis on the ``Marvel-ous Dwarfs'' zoom simulations \citep{Munshi21}, we find results consistent with the isolated galaxies in \Rom, where UDGs are more oblate. The median ($b/a$, $c/a$) values are (0.85, 0.29) for UDGs and (0.8, 0.56) for non-UDGs, though we note our zoom simulations contain only four UDGs. Since the ``Marvel-ous'' suite has the resolution to form cores, this suggests that core formation (or the lack thereof) is not influencing our shape analysis in \Rom. We note, however, that a recent study by \citet{KadoFong21} found that the inferred three-dimensional (3D) shapes of observed LSB galaxies are well characterized by oblate spheroids, with $c/a$ values higher than those of high-mass dwarfs with thick disks. The LSB galaxies in the \Rom\  simulations (defined as having $r_{\text{eff}}>1$ kpc and $\bar{\mu}_{\text{eff,g}}>24.3$ \sbu) were found to be in good agreement with the observations (see Figure 8 in \citet{KadoFong21}). 

This disparity in the isolated UDG and non-UDG morphologies suggests that the two populations have different formation histories. \citet{Wright21} found that isolated dwarfs in \Rom25 all undergo similar numbers of mergers, both major and minor, and that UDGs are primarily the products of early ($>8$~Gyr ago) major mergers. Here we consider what effect this has on the shape evolution of the isolated galaxies. Figure \ref{fig:Field-StellarOnlyShape} shows the stellar $b/a$ and $c/a$ axis ratios for the isolated galaxies as a function of the time elapsed since a galaxy's last major merger. A time of $t=0$ represents the time of a halo's final major merger, with a time of $t<0$ being the time preceding the final merger, and a time of $t>0$ being the time elapsed since the merger. The time $t=0$ is chosen to be the instant that the virial radii of the merging halos first overlap, following \citet{Hetznecker06}, and only time bins with at least five data points are plotted. The bold lines represent the populations' median values, while the shaded regions cover the 25$^{\text{th}}$ to 75$^{\text{th}}$ percentiles. Additionally, the galaxies are split into columns according to stellar mass. This binning, following \citet{Wright21}, separates our population into dependencies on UDG criteria, with the low-mass bin (149 galaxies) being largely dependent on effective radius only, the high-mass bin (298 galaxies) being dependent on central surface brightness only, and the intermediate-mass bin (224 galaxies) being dependent on both.  Note that all of the UDGs in the high-mass bin, and most of the other UDGs, undergo their last major merger (LMM) within the first 5 Gyr of the simulation, while the LMMs for non-UDGs are much more spread out in time. After their LMMs, both UDGs and non-UDGs experience a gradual increase in $b/a$. Since UDGs more typically have early LMMs, this could result in a $z=0$ $b/a$ disparity between dwarfs and UDGs, as seen in Figure \ref{fig:CvB}. However, when scaling to LMM time, we see that intermediate- and high-mass UDGs still evolve to higher $b/a$ values than even their early-merging non-UDG counterparts, suggesting that something in addition to early mergers is driving a change in shapes. The divergence from this trend in the lowest mass bin is likely a combination of overquenching due to resolution effects and these galaxies being genuinely dispersion-supported.

While UDGs are more likely to experience earlier LMMs, this does not fully explain the discrepancy in shapes.  This is consistent with \citet{Wright21}, who found that while isolated UDGs were largely influenced by early mergers, that alone was not sufficient to explain their formation. In fact, \citet{Wright21} found that the mergers  were not only earlier on average, but also tended to produce a larger increase in specific angular momentum. In Figure \ref{fig:Lambda}, we examine the evolution in halo spin, now also splitting the galaxy population based on early- and late-time mergers.  Here the spin parameter is the \citet{Bullock01} spin. Examining the early mergers (top row), we find that mergers that result in intermediate and massive UDGs at $z=0$ have more angular momentum, which results in a significant increase in halo spin beginning at the last major merger. Galaxies that experience LMMs also see an increase in spin, but this is much less likely to produce a UDG at $z=0$. We find that UDGs experience less of an increase in $c/a$ post-LMM; however, we note that the slight difference in the $c/a$ shape post-LMM may be attributed to the differences in spin of UDG and non-UDG LMMs. Thus, we confirm that the combination of early and high-spin mergers is required to form isolated UDGs, as demonstrated in \citet{Wright21}, and that this is the driving force behind the difference in shapes between isolated UDGs and isolated dwarfs. 
To summarize: we show that post-LMM, a dwarf galaxy's $b/a$ value increases as the galaxy relaxes, such that those with earlier mergers (UDGs) have larger $b/a$ values by $z=0$ (see Fig. \ref{fig:Field-StellarOnlyShape}).  For UDGs, $c/a$ remains lower for UDGs when compared to dwarf galaxies because their mergers have higher spin (see Figs \ref{fig:Field-StellarOnlyShape} and \ref{fig:Lambda}).  As such, we predict that observations of stellar shape can be used to identify the major merger history of dwarf galaxies and a formation channel for UDGs (early, high-spin mergers). 

\begin{table*}[th]

\begin{tabular}{cccccc}
\hline
Name & Surface Brightness (\sbu) & $r_{\text{eff}}$ (kpc) & S{\'e}rsic Index & Absolute Magnitude & Reference\\
\hline
G0-F & $\mu_{\text{g}}$(0) $> $24  & $r_{\text{eff}} >$ 1.5  & Free & N/A & \citet{Gband}\\
\hline
G0-1 & $\mu_{\text{g}}$(0) $> $24  & $r_{\text{eff}} >$ 1.5  & n = 1 & N/A & \citet{Gband}\\
\hline
R0 & $\mu_{\text{R}}$(0) $> $23.5  & $r_{\text{eff}} >$ 1.5  & Free & N/A & \citet{Forbes20}\\
\hline
RE & $\mu_{\text{eff,R}} > $23.5  & $r_{\text{eff}} >$ 1  & Free & $-12 > M_{\text{R}} > -16.5$ & \citet{Reff}\\
\hline
R{\=E} & 26.5 $> \langle\mu_{\text{R}}(r_{\text{eff}})\rangle >$ 24  & $r_{\text{eff}} >$ 1.5  & Free & N/A & \citet{SBavg}\\
\hline
\end{tabular}
\caption{\label{Definitions}A summary of the UDG definitions considered in This work.}
\end{table*}

\begin{table}[h]
\begin{center}
\begin{tabular}{ c|c|c c }
\hline
Definition & Environment & N$_{\text{UDG,faceon}}$ & Percentage\\
\hline
\multirow{3}{4em}{G0-F} & Isolated & 134 & $19.97\pm1.73$\\
                         & Satellite & 142 & $37.67\pm3.16$\\
                         & Cluster & 122 & $60.70\pm5.50$\\
\hline
\multirow{3}{4em}{G0-1} & Isolated & 241 & $35.92\pm2.31$\\
                         & Satellite & 230 & $61.01\pm4.02$\\
                         & Cluster & 167 & $83.08\pm6.43$\\
\hline
\multirow{3}{4em}{R0} & Isolated & 125 & $18.63\pm1.67$\\
                         & Satellite & 125 & $33.16\pm2.97$\\
                         & Cluster & 104 & $51.74\pm5.07$\\
\hline
\multirow{3}{4em}{RE} & Isolated & 454 & $67.66\pm3.18$\\
                         & Satellite & 309 & $81.96\pm4.66$\\
                         & Cluster & 168 & $83.58\pm6.45$\\
\hline
\multirow{3}{4em}{R\=E} & Isolated & 194 & $28.91\pm2.08$\\
                         & Satellite & 180 & $47.75\pm3.56$\\
                         & Cluster & 132 & $65.67\pm5.72$\\
\hline
\end{tabular}
\caption{The number of galaxies that are identified as UDGs at face-on orientation for each environment and UDG definition. The number is also given as a fraction of the environment's total dwarf population with Poisson error.}
\label{FaceonCount}
\end{center}
\end{table}

\section{Effects of Changing Definition and Orientation}
\label{sec:effects}

In Section \ref{sec:shapes}, we showed how the unique formation mechanism for isolated UDGs identified in \citet{Wright21} leads to differences in UDG and non-UDG morphology. During our study, we found that using alternate UDG criteria can cause these differences to disappear, removing any indication of the underlying formation mechanisms. To study the effect that the UDG definition has on both the above results and the resultant UDG population as a whole, we analyze our dwarf sample using multiple sets of UDG criteria.

\subsection{Definitions of UDGs}
\label{subsec:definitions}

To illustrate this point, we consider five different definitions of UDGs. Each of these definitions are used in the recent literature, and are summarized in Table \ref{Definitions}:
\begin{itemize}
    \item \textbf{G0-F}: this definition comes from \citet{Gband}. It requires the central $g$-band surface brightness to be fainter than 24 \sbu\ and the effective radius to be larger than 1.5 kpc. While performing the S\'ersic fits for this definition, we allow the S\'ersic index to vary as a free parameter. This is the definition outlined in Section \ref{subsec:UDG_ID} and used in the results presented in Section \ref{sec:shapes}.
    \item \textbf{G0-1}: this definition is the same as G0-F, but when performing the S\'ersic fits, the S\'ersic index is fixed at a value of 1, keeping in line with the original method in \citet{Gband}.
    \item \textbf{R0}: this definition comes from \citet{Forbes20}. It requires the central $R$-band surface brightness to be fainter than 23.5 \sbu\ and the effective radius to be larger than 1.5 kpc. While performing the S\'ersic fits for this definition, we allow the S\'ersic index to vary as a free parameter.
    \item \textbf{RE}: this definition comes from \citet{Reff}. It requires the effective $R$-band surface brightness to be fainter than 23.5 \sbu\ and the effective radius to be greater than 1 kpc. Additionally, the authors require the $R$-band absolute magnitude to be within -16.5 to -12. While performing the S\'ersic fits for this definition, we allow the S\'ersic index to vary as a free parameter. We note that the authors define the effective surface brightness as $L/(2\pi r^2_{\text{eff,R}}$) rather than $\mu_{\text{R}}$($r_{\text{eff}}$) (the value returned by the S\'ersic fit), so we adopt this method as well when identifying UDGs with this definition. Using $\mu_{\text{R}}$($r_{\text{eff}}$) results in a slight increase in the obtained value for $\mu_{\text{eff}}$ ($\approx1$ \sbu), but only results in a slight difference in the number of UDGs ($\approx0-4$\% of each environment's populations) and has no bearing on the results of this work.
    \item \textbf{R\=E}: this definition comes from \citet{SBavg}. It requires the average $R$-band surface brightness within the effective radius to be between 24-26.5 \sbu\ and the effective radius to be larger than 1.5 kpc. While performing the S\'ersic fits for this definition, we allow the S\'ersic index to vary as a free parameter.
\end{itemize}

\subsection{Definition Effects}
\label{subsec:def_effects}

The numbers of galaxies that are identified as UDGs are given in Table \ref{FaceonCount} for each environment and definition. The process of identification follows from Section \ref{subsec:UDG_ID} for each definition  (i.e., we first orient each galaxy to a face-on position, before exploring orientation effects below), with the appropriate size and surface brightness restrictions applied. Regardless of definition, the percentage of dwarf galaxies that would be identified as UDGs generally increases with the density of the environment. This trend with environment matches the predictions from \citet{Martin19} using the Horizon-AGN simulation \citep{Dubois14}. The authors found that UDGs (here defined as $\langle\mu_{\text{R}}(r_{\text{eff}})\rangle>24.5$) represent a significant percentage of the galaxy population, and this percentage increases with environmental density. These results also agree with those from \citet{Jackson20}, who found that UDGs exist in large numbers in groups and the field within the NewHorizon Simulation \citep{Dubois21}, as well as agreeing with observational results \citep[e.g.][]{vanderBurg17}.

Although the trend with environment is ubiquitous across most definitions, the total number of UDGs identified is not. The G0-F and R0 definitions are much more restrictive in their UDG identification compared to the RE definition, which identifies the largest population of UDGs in all environments. The G0-1 and R\=E definitions tend to fill in the middle ground between the G0-F/R0 and RE definitions in terms of the size of the UDG populations.

\subsection{Orientation Effects}
\label{subsec:rotation}

\begin{figure*}
\begin{subfigure}{\linewidth}
  \centering
  \includegraphics[width=\linewidth]{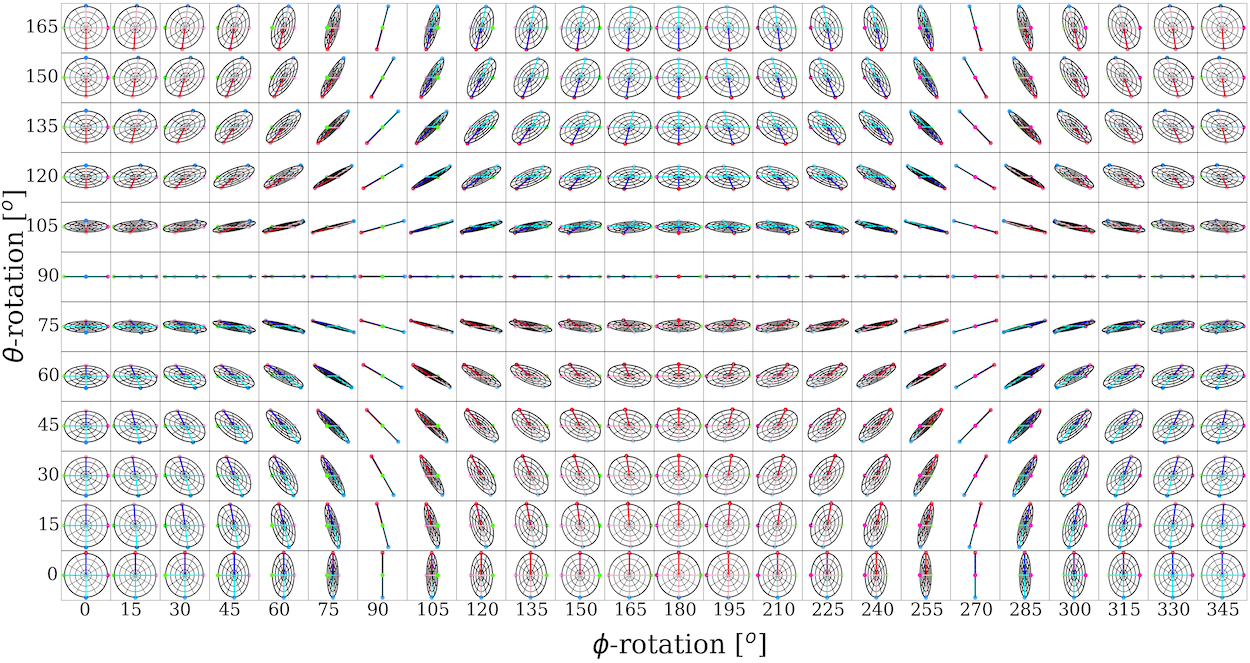}
  \caption{}
  \label{subfig:RotationGuide}
\end{subfigure}
\newline
\begin{subfigure}{\linewidth}
  \centering
  \includegraphics[width=\linewidth]{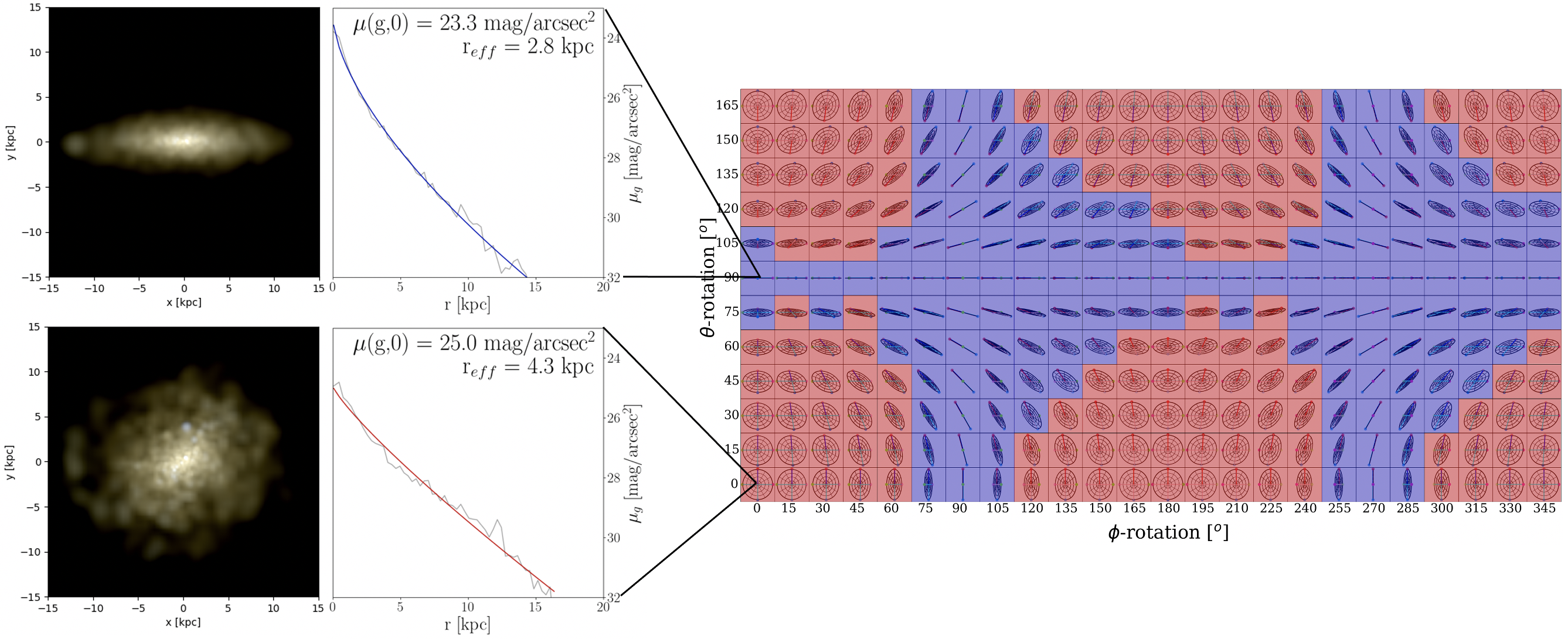}  
  \caption{}
  \label{subfig:RotationExample}
\end{subfigure}
\caption{(a) A visual representation of the rotated galaxy positions considered in this work. The sample `galaxy' is a flat, circular disk with one face blue and the opposite face red. The disk has a single dark-colored radius and multicolored vertices to help visualize the rotations. (b) \textbf{Right:} an example of the rotation-dependent UDG status of an isolated galaxy from \Rom25 with $M_\star = 3.26\times10^8$ $M_\odot$. The grid shows whether the galaxy is identified as a UDG (red) or a standard dwarf (blue) at the given rotation under the G0-F definition. \textbf{Left:} example $g$-band surface brightness profiles and their associated S\'ersic fits of the isolated galaxy for ($\theta,\phi$) rotations of ($0\degree,0\degree$; i.e., face-on) and ($90\degree,0\degree$; i.e., edge-on). The fit parameters are given in the upper right-hand corners of the surface brightness plots. To the left of the plots are UVI images of the galaxy at the two orientations showing all features brighter than 32 \sbu (the region being fit by the S\'ersic profile).}
\label{fig:HaloGuide}
\end{figure*}

Since all of our definitions of UDGs are derived from surface brightness profiles, a galaxy's status as a UDG is intrinsically dependent on its viewing orientation. It is standard practice when searching for UDGs within simulations to first orient the galaxy to a face-on position \citep{Reff,Tremmel20,Wright21}. This, in theory, will maximize the galaxy's effective radius while minimizing its central surface brightness, optimizing its likelihood of being identified as a UDG. Here, we study whether this assumption holds true for all galaxies, and how a galaxy's orientation could affect its status as a UDG. Some previous studies \citep[e.g.][]{Chan18,Jackson20,CardonaBarrero20} have considered multiple viewing angles (such as the three primary axes or a span of inclinations), but here we consider 288 positions from incremental rotations over two axes perpendicular to the line of sight (see Figure \ref{subfig:RotationExample} for a visual guide). These positions, in their entirety, detail how the galaxy is viewed from any position in 3D space.

As we describe in Section \ref{subsec:UDG_ID}, when analyzing a galaxy, we first orient it to a face-on position. After the initial orientation, the galaxy is then rotated by an angle $\theta$ around an axis perpendicular to the line of sight (the angular momentum vector in the face-on position). Next, the galaxy is rotated by an angle $\phi$ around the axis perpendicular to the line of sight and previous axis of rotation. An azimuthally averaged surface brightness profile for the galaxy is generated and fit with a S\'ersic profile that is used to determine whether the galaxy is identified as a UDG at the given orientation. The $\theta$-rotations and $\phi$-rotations considered in this work span the spaces of [0\degree,180\degree) and [0\degree,360\degree), respectively, both in increments of 15\degree. Since the radial surface brightness profile of a galaxy is unaltered under rotations around the line of sight, any ($\theta,\phi$) where $\theta<180$\degree\ is analogous to a rotation of ($180\degree+\theta,180\degree-\phi\ [mod360]$) where $[mod360]$ (modulo 360\degree) means that if $180\degree-\phi=-x<0$, the value becomes $360\degree-x$. Thus, the $\theta$ ($\phi$) rotation space of [180\degree,360\degree) ([0\degree,360\degree)) is bijectively mapped to the [0\degree,180\degree) ([0\degree,360\degree)) space, and we have analyzed all unique positions. The top panel of Figure \ref{subfig:RotationGuide} illustrates the rotations considered in this work using a flat, circular disk.

The numbers of galaxies that are identified as UDGs when considering any orientation are given in Table \ref{AnyCount} for each environment and definition. The percentages of dwarf galaxies that are UDGs still increase with environmental density, but the numbers of UDGs have increased by a significant margin in almost every case. The value $\Delta$ in Table \ref{AnyCount} represents the number of UDGs gained when considering orientations other than face-on, i.e. the number of galaxies that are \textit{not} classified as a UDGs at a face-on orientation but \textit{are} classified as such at some other orientation. Although this value decreases with environmental density, it is present in all definitions, indicating that by analyzing all galaxies in a face-on position, one is likely to omit a significant fraction of the galaxies that could potentially be identified as UDGs. This fraction can be quite large in the isolated environment, ranging from 7-20\% of isolated dwarf galaxies, but is more minor in the denser environments, ranging from 2-17\% in satellites and 0-13\% in the cluster. This also hints at isolated galaxies' classifications as UDGs being more sensitive to orientation, a result we would expect, based on their more ``disky'' morphology, seen in Figure \ref{fig:CvB}.

\begin{table}
\begin{center}
\begin{tabular}{c|c|c c|c}
\hline
Def.& Env. & N$_{\text{UDG,any}}$ & Percentage & $\Delta_{\mathrm{face-on}}$ (\%)\\
\hline
\multirow{3}{4em}{G0-F} & Isolated & 191 & $28.5\pm2.1$ & 29.5\\
                         & Satellite & 186 & $49.3\pm3.6$ & 23.7\\ 
                         & Cluster & 148 & $73.6\pm6.1$ & 17.6\\ 
\hline
\multirow{3}{4em}{G0-1} & Isolated & 379 & $56.5\pm2.9$ & 36.4\\
                         & Satellite & 293 & $77.7\pm4.5$ & 21.5\\
                         & Cluster & 186 & $92.5\pm6.8$ & 10.2\\
\hline
\multirow{3}{4em}{R0} & Isolated & 197 & $29.4\pm2.1$ & 36.5\\
                         & Satellite & 176 & $46.7\pm3.5$ & 28.0\\
                         & Cluster & 129 & $64.2\pm5.7$ & 19.4\\
\hline
\multirow{3}{4em}{RE} & Isolated & 502 & $74.8\pm3.3$ & 10.0\\
                         & Satellite & 320 & $84.9\pm4.7$ & 3.4\\
                         & Cluster & 168 & $83.6\pm6.5$ & 0.0\\
\hline
\multirow{3}{4em}{R\=E} & Isolated & 260 & $38.8\pm2.4$ & 25.4\\
                         & Satellite & 224 & $59.4\pm4.0$ & 19.6\\
                         & Cluster & 155 & $77.1\pm6.2$ & 14.8\\
\hline
\end{tabular}
\caption{The number of galaxies that are identified as UDGs at any orientation for each environment and UDG definition. The number is also given as a fraction of the environment's total dwarf population with Poisson error. The value $\Delta_{\mathrm{face-on}}$ is the fraction of UDGs that were not identified using the face-on orientation, as used in Table \ref{FaceonCount}.}
\label{AnyCount}
\end{center}
\end{table}

\begin{figure*}
\centering
\includegraphics[width=\linewidth]{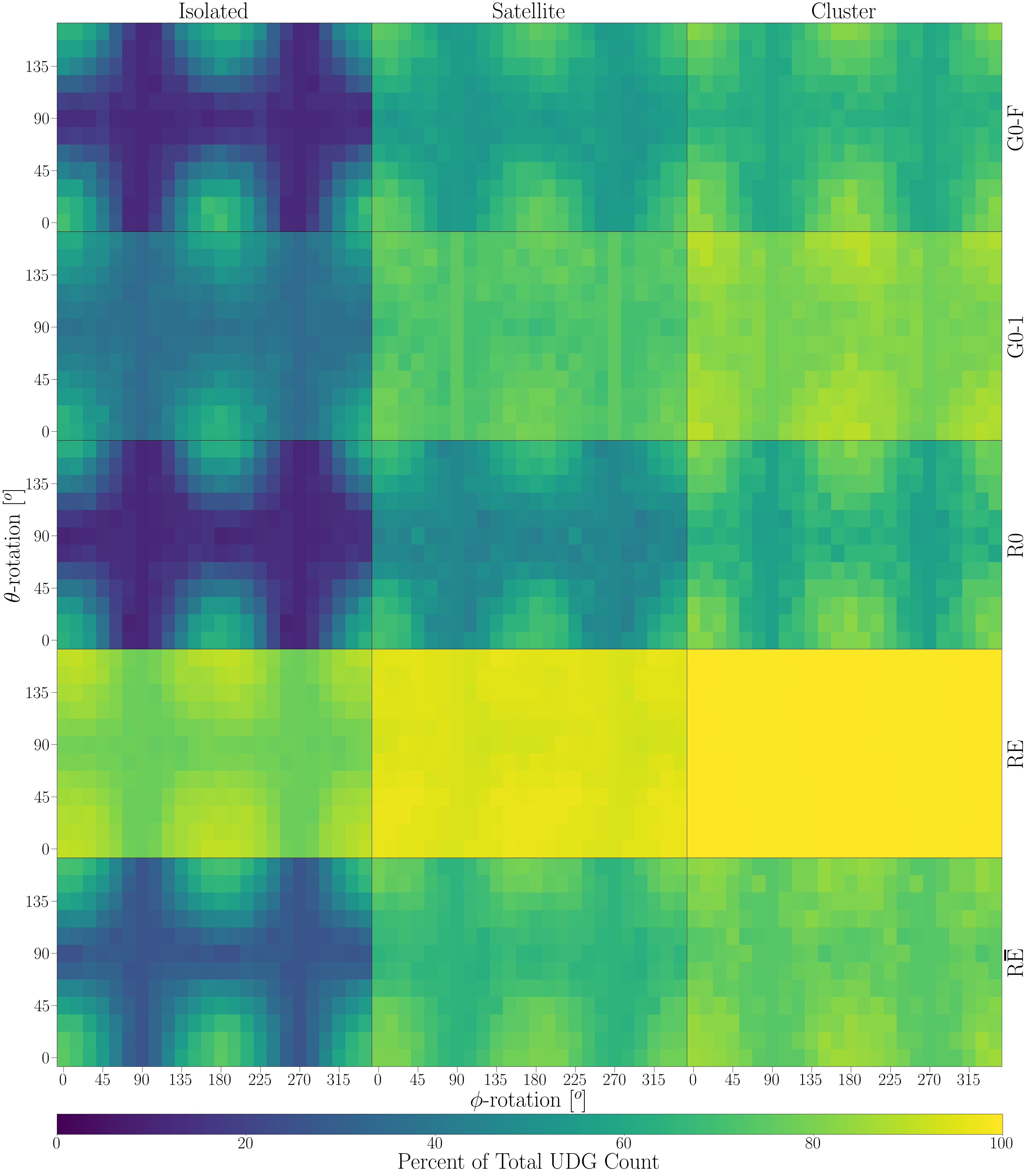}
\caption{Color grids representing the UDG populations for each environment and definition of UDG. Each point on the grid represents the UDG population at that particular orientation. The value plotted is the number of galaxies that are identified as UDGs at that orientation as a percentage of the total number of galaxies that are identified as UDGs at any orientation (i.e., the ``Total UDG Count'' for that environment and definition). In all environments and definitions, the UDG count increases as you move from edge-on positions to face-on ones. In each definition, there is the most rotation dependence (color contrast) in the isolated environment and the least in the cluster environment.}
\label{fig:MasterAll}
\end{figure*}

To further investigate whether isolated UDGs are less robust to orientation, we look at the different environments' total UDG populations at each specific orientation. Figure \ref{fig:MasterAll} summarizes these results with a color grid for each environment and UDG definition. Each point on the grid represents the total number of galaxies that are identified as ultra-diffuse at that orientation as a fraction of the total number of UDGs in that environment (N$_{\text{UDG,any}}$ from Table \ref{AnyCount}). In all definitions, the isolated UDG populations demonstrate much stronger orientation dependence than their denser-environment counterparts. The more restrictive definitions, G0-F and R0, show the strongest orientation dependence, with the percentages varying by $\sim 55-60$\% in the isolated environment. In contrast, the least restrictive definition, RE, also appears to be the least orientation-dependent, with the percentages varying by $\sim 15$\% in the isolated environment and essentially 0\% in the cluster environment (there are only two cluster galaxies that are identified as RE-UDGs at some orientations, but not all). The RE definition, in general, seems to be identifying the highest fraction of dwarf galaxies as UDGs at all orientations, which is expected from its $\Delta$ values in Table \ref{AnyCount} being the smallest (especially when considering that the RE N$_{\text{UDG,any}}$ values were typically larger than those for other definitions). Again, the G0-1 and R\=E definitions seem to fill out the middle ground, both in terms of orientation dependence and the average percentage of UDGs retained across orientations. In all cases, the UDG populations are maximized in the positions that are more face-on than edge-on. This suggests that if simulators were to analyze all galaxies from the same orientation in order to identify UDGs, choosing face-on would likely produce the largest UDG population, but we have shown that allowing for any orientation will produce the maximum possible number of UDGs.

\begin{figure*}[t]
\centering
\includegraphics[width=\linewidth]{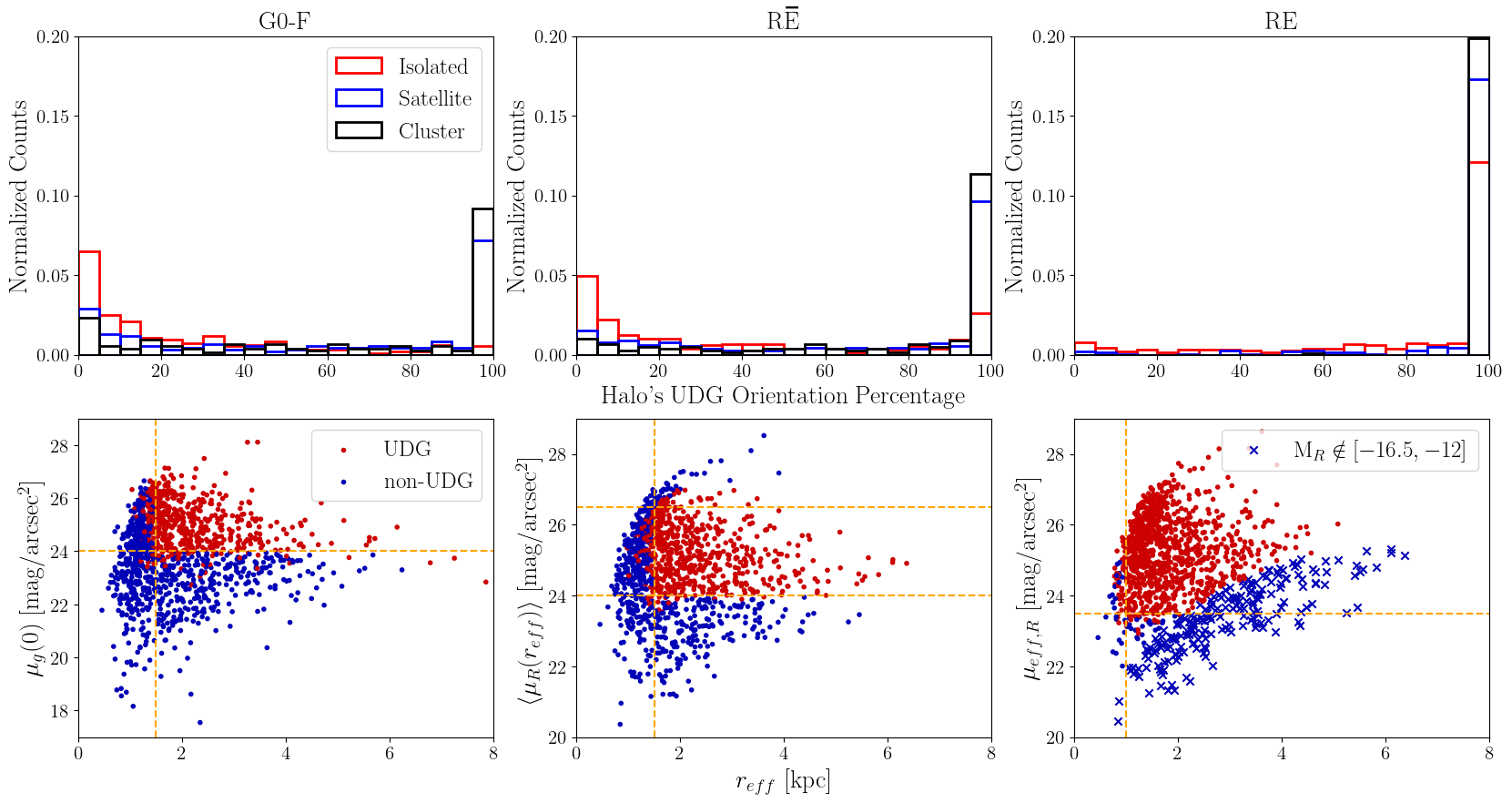}
\caption{
\textbf{Top: }normalized histograms of what percentage of the time a galaxy would be identified as a UDG when oriented randomly. Galaxies that are never considered ultra-diffuse are not included. From left to right, the UDG definitions used were G0-F, R\=E, and RE. In the G0-F definition (the most restrictive), a large number of galaxies in all environments exist in the 0-20\% bins, indicating a high rotation dependence. In the RE definition (the least restrictive), most galaxies exist in the 95-100\% bin, indicating little to no rotation dependence.
\textbf{Bottom: }surface brightness versus effective radii plots, including dotted lines marking the selection criteria for UDGs. Galaxies are plotted using their face-on values. Points labeled as UDG (red) are those that are identified as such at any orientation, while points labeled as non-UDG (blue) are never classified as such. As above, the UDG definitions used were G0-F, R\=E, and RE. The blue ``X''s denote galaxies that violate the $R$-band absolute magnitude restriction under the RE definition. Different definitions identify different subsections of the dwarf population to be UDGs, with the RE definition selecting a notably larger area.
}
\label{fig:RvMu}
\end{figure*}

In addition to studying how the statistics of UDG populations vary with orientation, we analyze how the individual galaxies vary under rotation with each definition, as well as how the definitions change the populations of galaxies identified as ultra-diffuse. Figure \ref{fig:RvMu} shows this analysis for the G0-F, R\=E, and RE definitions (the R0 and G0-1 definitions are approximated here by the G0-F and R\=E definitions, respectively). The top row shows the normalized distributions of individual galaxies' UDG percentages, i.e., what percentage of the time a galaxy would be identified as a UDG if its orientation were drawn at random (cos($\theta$) uniform in [-1,1] and $\phi$ uniform in [0,2$\pi$]). In the G0-F and R\=E definitions, the isolated galaxies tend to occupy the low-percentage space, particularly 0-15\%, while a large fraction of satellite and cluster galaxies occupy the 95-100\% bin. The R\=E definition shows very similar results to the G0-F definition, but sees some galaxies in all environments shift from the lower percentages to the higher ones. In contrast, the RE definition sees most galaxies in all environments occupying the 95-100\% bin, mirroring the lack of orientation dependence seen in Figure \ref{fig:MasterAll}. This implies that if a galaxy is identified as a UDG by the RE definition at some orientation, it is likely to be identified as such at any orientation. 

The bottom row shows all of our dwarfs in surface brightness versus effective radius space with their face-on values. Also shown are the boundaries for UDG classification. The RE definition, while being the most robust to orientation,  demarcates the largest area of our dwarf population as ultra-diffuse. A large number of galaxies are not identified as RE-UDGs only because they violate the absolute magnitude restriction.

\begin{figure}
\centering
\includegraphics[width=\linewidth]{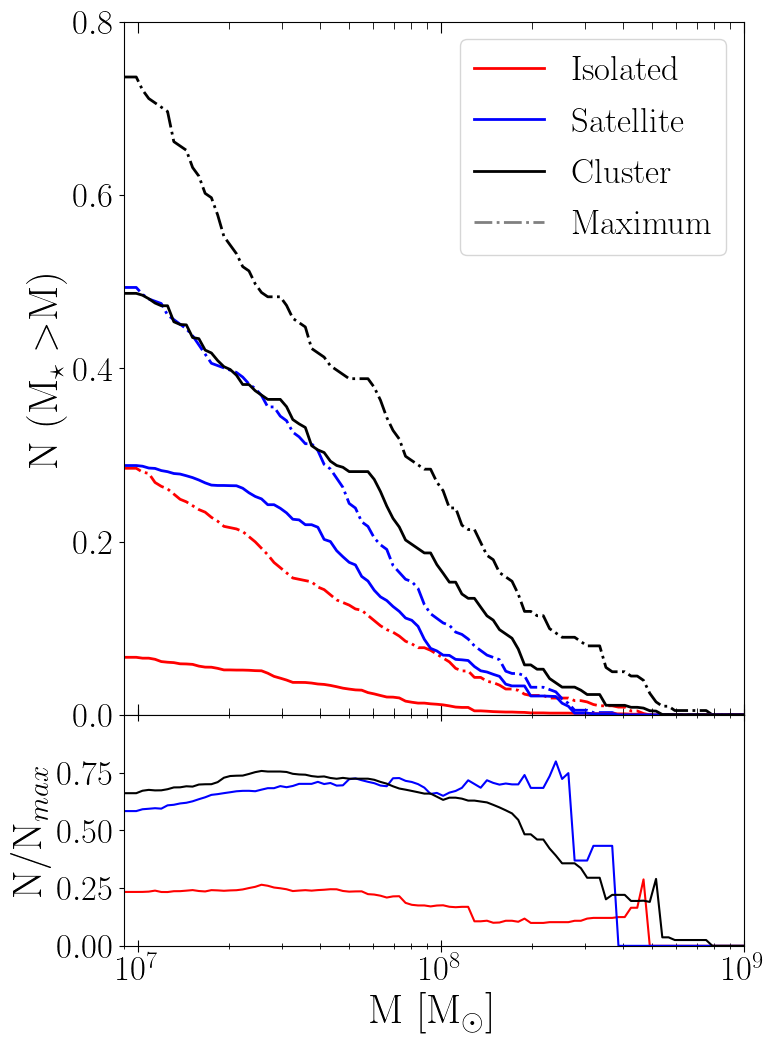}
\caption{
\textbf{Top: }the mass functions for G0-F-UDGs when all dwarfs are oriented randomly, normalized to the size of the environment's dwarf population. The dotted lines are the maximum situation where every galaxy that could be identified as a UDG is oriented as such.
\textbf{Bottom: }the UDG mass functions for randomly oriented galaxies divided by the maximum number mass function for each environment. The difference between the maximum and random orientation functions is largest in the isolated environment, where the UDGs are most rotation-dependent.
}
\label{fig:MassFcn}
\end{figure}

Using the individual galaxies' UDG percentages, we can construct mass functions. Figure \ref{fig:MassFcn} shows the UDG mass functions for each environment using the G0-F definition, assuming each galaxy is oriented randomly. Rather than using the binary choice of UDG or non-UDG (adding a 1 or 0 to the mass function), a galaxy's contribution is determined by its UDG percentage (the value plotted in the top row of Figure \ref{fig:RvMu}) and the resultant mass function is normalized to the size of the environment's dwarf population. The dotted lines are the `maximum' mass functions, representing the idealized situation where every galaxy that could be identified as a UDG at some orientation is counted. The bottom plot shows the fraction of the maximum mass function that is occupied by the random orientation mass function. In the isolated and cluster environments, the $N/N_{\text{max}}$ percentage is relatively constant between 10$^{7}$ and $10^8$ $M_\odot$, after which it starts to decrease. The satellite fraction steadily increases until $3\times 10^8$ $M_\odot$ where the sample size approaches zero. When switching from the maximum mass function to one that accounts for random orientations, we see the greatest difference in the isolated UDGs, as demonstrated in the bottom panel of Figure \ref{fig:MassFcn}.  This is consistent with our shape results: compared to the cluster and satellite UDGs, isolated UDGs have a  less spherical shape (more oblate), which implies that orientation will alter their classification.  The more spherical populations are more robust to this, again as demonstrated in the bottom panel of Figure \ref{fig:MassFcn}. None of the environments yield perfectly spherical populations, thus all populations have mass functions that depend on rotation. This is important to consider when comparing simulations to observations. Since simulations typically assemble their UDG populations from face-on analysis, it may not be appropriate to compare the resultant UDG mass function to one from observations, where galaxies are oriented randomly.

Recognizing that 3D shape may not be readily apparent in observations, we have also constructed ellipticity functions for all of our galaxies at random orientations. For each galaxy, a mock ellipsoid with the galaxy's $b/a$ and $c/a$ axis ratios is generated at a randomly selected orientation. The ellipsoid is then projected into a two-dimensional (2D) plane, and the projected ellipticity (1-$b/a$) (where $a$ and $b$ are now the projected major and minor axis ratios, respectively) is measured for the resultant ellipse traced out by the edge of the projection.

\begin{figure*}
\centering
\includegraphics[width=\linewidth]{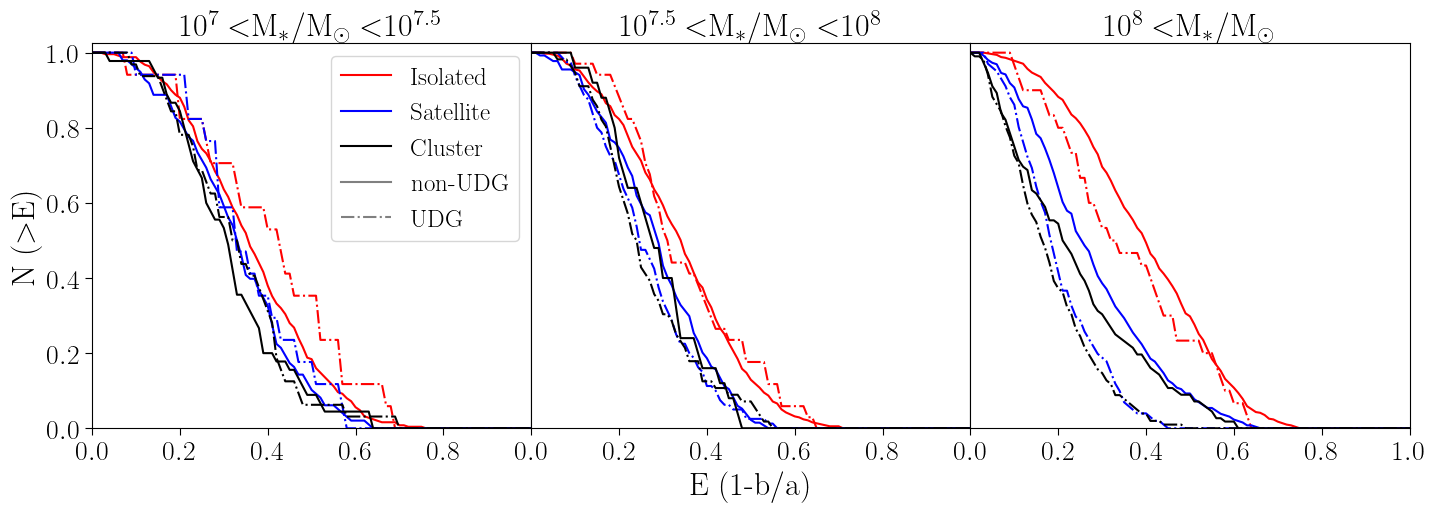}
\caption{Normalized ellipticity functions for 2D projections of our galaxies at random orientations. The trends shown are the median values after 100 iterations of random orientations. We do not expect that measurements of 2D ellipticity will show the shape disparity between isolated UDGs and non-UDGs.}
\label{fig:EllFcn}
\end{figure*}

Figure \ref{fig:EllFcn} shows the median normalized ellipticity functions for all of our galaxies for 100 random orientations. The sample is split into mass bins (as with Figures \ref{fig:Field-StellarOnlyShape} and \ref{fig:Lambda}), and a galaxy's status as UDG is determined at that particular random orientation. In the intermediate- and high-mass bins, the difference in the distributions of isolated UDGs and non-UDGs is quite small, though the direction of the trends is the same as in our 3D results. The combination of our previously discussed orientation effects, along with the projection from 3D to 2D erase some of the shape difference we see in Section \ref{sec:shapes}. We conclude that it may be difficult to use measurements of 2D ellipticity to infer a difference in isolated UDG and non-UDG morphologies. However, methods for obtaining 3D shapes, akin to those in \citet{KadoFong21}, are promising avenues for inferring 3D shapes and thus confirming the shape differences we predict.

\begin{figure}
  \centering
  \includegraphics[width=\linewidth]{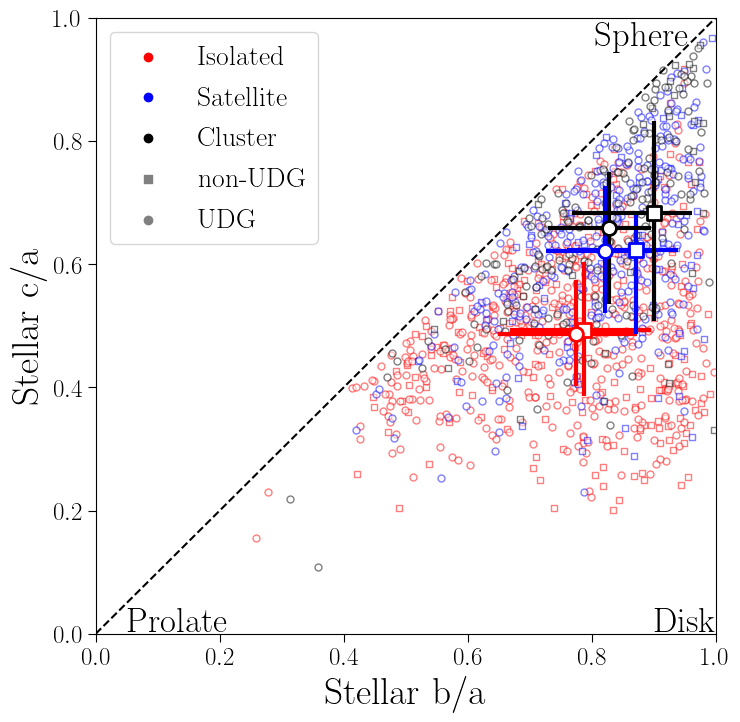}
\caption{The stellar $c/a$ axis ratios plotted against the $b/a$ axis ratios for the galaxies' stellar distributions at $z=0$ in all environments. The populations are separated into UDGs and non-UDGs under the RE definition. The bold points represent the medians of the populations, while the error bars show the 25$^{\text{th}}$ to 75$^{\text{th}}$ percentiles. Under the RE definition, the disparity in isolated UDG and non-UDG shapes seen in Figure \ref{fig:CvB} is no longer present.}
\label{fig:CvB_RE}
\end{figure}

\subsection{Effects on Previous Results}
\label{subsec:shapeeffects}

We have shown that altering the criteria for UDGs and accounting for different orientations can have a large effect on the resultant UDG population. This effect means that various groups studying UDGs under varying definitions could potentially be looking at vastly different populations. This could make reaching a cohesive understanding of UDGs (including formation mechanisms and whether they exclusively reside in dwarf dark matter halos) difficult, if not impossible. 

As an example: in Section \ref{sec:shapes}, we found a disparity between isolated UDG and non-UDG morphologies that, after investigation, revealed the formation mechanism of early, high-spin mergers. This disparity, highlighted in Figure \ref{fig:CvB}, was present when identifying UDGs according to the G0-F definition. However, recreating Figure \ref{fig:CvB} under the different definitions can provide drastically different results. Figure \ref{fig:CvB_RE} shows the $c/a$ versus $b/a$ axis ratios for the dwarf halos with UDGs identified via the RE definition. We see now that the isolated galaxies show no disparity between UDGs and non-UDGs, with their medians and percentiles lying nearly on top of one another. Together, Figures \ref{fig:RvMu} \& \ref{fig:CvB_RE} indicate that the RE definition is permissive to the extent that the UDG and non-UDG populations are effectively homogenized, thus no underlying formation mechanisms or physics would be unique to the UDG population.

\section{Conclusions}
\label{sec:conclusions}

We have selected 1249 dwarf galaxies in isolated, satellite, and cluster environments from the \Rom25 and \Rom C simulations. We analyze the shapes of these galaxies as well as the resultant UDG population under five definitions and 288 orientations.
Our results can be summarized as follows:
\begin{itemize}
    \item A morphological disparity exists between isolated UDGs' and non-UDGs' stellar distributions, with UDGs having notably larger $b/a$ and smaller $c/a$ axis ratios, both differing by $\approx 0.1$ (see Figure \ref{fig:CvB}). This more oblate-triaxial morphology is in agreement with \citet{Rong20} and \citet{KadoFong21}.
    \item Analysis of the isolated galaxies' morphological histories shows that UDGs are primarily the products of early mergers (see Figure \ref{fig:Field-StellarOnlyShape}). Further investigation shows that early mergers alone are not sufficient; high spins are also required (see Figure \ref{fig:Lambda}).
    \item Changing the criteria for being ultra-diffuse can result in largely different UDG populations, with the percentage of dwarfs that are identified as UDGs ranging from 19-65\%, 33-70\%, and 50-83\% in the isolated, satellite, and cluster environments, respectively (see Table \ref{FaceonCount}).
    \item A galaxy's status as a UDG is dependent on its orientation, and this dependence is strongest in the isolated environment (see Figures \ref{fig:HaloGuide} \& \ref{fig:MasterAll}). Additionally, more restrictive definitions (like G0-F and R0) result in a stronger orientation dependence (see Figures \ref{fig:MasterAll} \& \ref{fig:RvMu}). This orientation dependence also manifests when comparing the mass functions of all potential UDGs to those identified from a random orientation (see Figure \ref{fig:MassFcn}).
    \item Although it is standard practice to identify UDGs from a face-on orientation, we find a significant number of galaxies that are identified as UDGs at some orientation, but do not meet the criteria when face-on. Depending on the UDG definition, there are 48-138, 8-63, and 0-26 such galaxies in isolated, satellite, and cluster environments, respectively (see Table \ref{AnyCount}).
    \item Changing the definition of a UDG can change or obscure the underlying physics of the resultant UDG population. The unique formation mechanism of early, high-spin mergers in isolated UDGs under the restrictive definitions (like G0-F and R0) is not highlighted under the less restrictive RE definition, where the UDG and non-UDG populations are effectively homogenized (see Figures \ref{fig:CvB} \& \ref{fig:CvB_RE}). 
\end{itemize}

We have shown that the population of dwarfs that are considered UDGs, and whether that population exhibits a morphological distinction or formation channel from non-UDGs, is drastically dependent on choice of definition. With all of the outstanding questions around UDG formation, it is clear that different conclusions can be reached when identifying separate subsets of dwarfs as ``ultra-diffuse''. In order for more advanced questions (like those of formation and evolution) to be answered cohesively, we must first come to a cohesive methodology for identifying UDGs. More restrictive definitions (like G0-F and R0) seem to identify a subset of isolated dwarf galaxies with unique shape histories, indicating a specific channel of formation through early, high-spin mergers, while less restrictive definitions identify such a large population of UDGs that the UDG and non-UDG populations are roughly identical. In short, it seems that whether UDGs are a true physical phenomenon or simply a product of definition depends on the definition itself.

\acknowledgments
F.D.M. and J.D.V. acknowledge support from the University of Oklahoma. Long before the University of Oklahoma was established, the land on which the University now resides was the traditional home of the “Hasinais” Caddo Nation and Kirikiris Wichita and Affiliated Tribes. F.D.M. and J.D.V. acknowledge this territory once also served as a hunting ground, trade exchange point, and migration route for the Apache, Comanche, Kiowa, and Osage nations. Today, 39 tribal nations dwell in the state of Oklahoma as a result of settler and colonial policies that were designed to assimilate Native people. 
MT is supported by an NSF Astronomy and Astrophysics Postdoctoral Fellowship under award AST-2001810. The Romulus simulations are part of the Blue Waters sustained-petascale computing project, which is supported by the National Science Foundation (awards OCI-0725070 and ACI-1238993) and the state of Illinois. Blue Waters is a joint effort of the University of Illinois at Urbana–Champaign and its National Center for Supercomputing Applications. This work is also part 12 of a Petascale Computing Resource Allocations allocation supported by the National Science Foundation (award number OAC-1613674). This work also used the Extreme Science and Engineering Discovery Environment (XSEDE), which is supported by National Science Foundation grant number ACI-1548562.
This work was partially performed at the Aspen Center for Physics, which is supported by National Science Foundation grant PHY-1607611.

The authors thank Jenny Greene and Erin Kado-Fong for useful discussions related to this manuscript.  The authors also thank Nicole Sanchez for use of an HECC allocation.
The python packages \textsc{matplotlib} \citep{Hunter07}, \textsc{numpy} \citep{vanderWalt11}, \textsc{pynbody} \citep{pynbody}, and \textsc{tangos} \citep{Tangos} were used in the analysis.\\

\bibliography{bibliography}{}

\begin{thebibliography}{}
\expandafter\ifx\csname natexlab\endcsname\relax\def\natexlab#1{#1}\fi
\providecommand{\url}[1]{\href{#1}{#1}}
\providecommand{\dodoi}[1]{doi:~\href{http://doi.org/#1}{\nolinkurl{#1}}}
\providecommand{\doeprint}[1]{\href{http://ascl.net/#1}{\nolinkurl{http://ascl.net/#1}}}
\providecommand{\doarXiv}[1]{\href{https://arxiv.org/abs/#1}{\nolinkurl{https://arxiv.org/abs/#1}}}

\bibitem[{{Abel} {et~al.}(1997){Abel}, {Anninos}, {Zhang}, \&
  {Norman}}]{Abel97}
{Abel}, T., {Anninos}, P., {Zhang}, Y., \& {Norman}, M.~L. 1997, \na, 2, 181,
  \dodoi{10.1016/S1384-1076(97)00010-9}

\bibitem[{{Amorisco} \& {Loeb}(2016)}]{Amorisco16}
{Amorisco}, N.~C., \& {Loeb}, A. 2016, \mnras, 459, L51,
  \dodoi{10.1093/mnrasl/slw055}

\bibitem[{{Amorisco} {et~al.}(2018){Amorisco}, {Monachesi}, {Agnello}, \&
  {White}}]{Amorisco18}
{Amorisco}, N.~C., {Monachesi}, A., {Agnello}, A., \& {White}, S.~D.~M. 2018,
  \mnras, 475, 4235, \dodoi{10.1093/mnras/sty116}

\bibitem[{{Bellovary} {et~al.}(2010){Bellovary}, {Governato}, {Quinn},
  {Wadsley}, {Shen}, \& {Volonteri}}]{Bellovary10}
{Bellovary}, J.~M., {Governato}, F., {Quinn}, T.~R., {et~al.} 2010, \apjl, 721,
  L148, \dodoi{10.1088/2041-8205/721/2/L148}

\bibitem[{{Black}(1981)}]{Black81}
{Black}, J.~H. 1981, \mnras, 197, 553, \dodoi{10.1093/mnras/197.3.553}

\bibitem[{{Borlaff} {et~al.}(2019){Borlaff}, {Trujillo}, {Rom{\'a}n},
  {Beckman}, {Eliche-Moral}, {Infante-S{\'a}inz}, {Lumbreras-Calle}, {de
  Almagro}, {G{\'o}mez-Guijarro}, {Cebri{\'a}n}, {Dorta}, {Cardiel},
  {Akhlaghi}, \& {Mart{\'\i}nez-Lombilla}}]{Borlaff19}
{Borlaff}, A., {Trujillo}, I., {Rom{\'a}n}, J., {et~al.} 2019, \aap, 621, A133,
  \dodoi{10.1051/0004-6361/201834312}

\bibitem[{{Bromm} {et~al.}(2001){Bromm}, {Ferrara}, {Coppi}, \&
  {Larson}}]{Bromm01}
{Bromm}, V., {Ferrara}, A., {Coppi}, P.~S., \& {Larson}, R.~B. 2001, \mnras,
  328, 969, \dodoi{10.1046/j.1365-8711.2001.04915.x}

\bibitem[{{Bryan} \& {Norman}(1998)}]{BryanNorman98}
{Bryan}, G.~L., \& {Norman}, M.~L. 1998, \apj, 495, 80, \dodoi{10.1086/305262}

\bibitem[{{Bullock} {et~al.}(2001){Bullock}, {Dekel}, {Kolatt}, {Kravtsov},
  {Klypin}, {Porciani}, \& {Primack}}]{Bullock01}
{Bullock}, J.~S., {Dekel}, A., {Kolatt}, T.~S., {et~al.} 2001, \apj, 555, 240,
  \dodoi{10.1086/321477}

\bibitem[{{Burkert}(2017)}]{Burkert17}
{Burkert}, A. 2017, \apj, 838, 93, \dodoi{10.3847/1538-4357/aa671c}

\bibitem[{{Butsky} {et~al.}(2019){Butsky}, {Burchett}, {Nagai}, {Tremmel},
  {Quinn}, \& {Werk}}]{Butsky19}
{Butsky}, I.~S., {Burchett}, J.~N., {Nagai}, D., {et~al.} 2019, \mnras, 490,
  4292, \dodoi{10.1093/mnras/stz2859}

\bibitem[{{Cardona-Barrero} {et~al.}(2020){Cardona-Barrero}, {Di Cintio},
  {Brook}, {Ruiz-Lara}, {Beasley}, {Falc{\'o}n-Barroso}, \&
  {Macci{\`o}}}]{CardonaBarrero20}
{Cardona-Barrero}, S., {Di Cintio}, A., {Brook}, C. B.~A., {et~al.} 2020,
  \mnras, 497, 4282, \dodoi{10.1093/mnras/staa2094}

\bibitem[{{Carleton} {et~al.}(2019){Carleton}, {Errani}, {Cooper},
  {Kaplinghat}, {Pe{\~n}arrubia}, \& {Guo}}]{Carleton19}
{Carleton}, T., {Errani}, R., {Cooper}, M., {et~al.} 2019, \mnras, 485, 382,
  \dodoi{10.1093/mnras/stz383}

\bibitem[{{Carleton} {et~al.}(2021){Carleton}, {Guo}, {Munshi}, {Tremmel}, \&
  {Wright}}]{Carleton21}
{Carleton}, T., {Guo}, Y., {Munshi}, F., {Tremmel}, M., \& {Wright}, A. 2021,
  \mnras, 502, 398, \dodoi{10.1093/mnras/stab031}

\bibitem[{{Cen}(1992)}]{Cen92}
{Cen}, R. 1992, \apjs, 78, 341, \dodoi{10.1086/191630}

\bibitem[{{Chamba} {et~al.}(2020){Chamba}, {Trujillo}, \& {Knapen}}]{Chamba20}
{Chamba}, N., {Trujillo}, I., \& {Knapen}, J.~H. 2020, in Contributions to the
  XIV.0 Scientific Meeting (virtual) of the Spanish Astronomical Society, 20

\bibitem[{{Chan} {et~al.}(2018){Chan}, {Kere{\v{s}}}, {Wetzel}, {Hopkins},
  {Faucher-Gigu{\`e}re}, {El-Badry}, {Garrison-Kimmel}, \&
  {Boylan-Kolchin}}]{Chan18}
{Chan}, T.~K., {Kere{\v{s}}}, D., {Wetzel}, A., {et~al.} 2018, \mnras, 478,
  906, \dodoi{10.1093/mnras/sty1153}

\bibitem[{{Christensen} {et~al.}(2014){Christensen}, {Governato}, {Quinn},
  {Brooks}, {Shen}, {McCleary}, {Fisher}, \& {Wadsley}}]{Christensen14}
{Christensen}, C.~R., {Governato}, F., {Quinn}, T., {et~al.} 2014, \mnras, 440,
  2843, \dodoi{10.1093/mnras/stu399}

\bibitem[{{Conselice}(2018)}]{Conselice18}
{Conselice}, C.~J. 2018, Research Notes of the American Astronomical Society,
  2, 43, \dodoi{10.3847/2515-5172/aab7f6}

\bibitem[{{Dalcanton} {et~al.}(1997){Dalcanton}, {Spergel}, {Gunn}, {Schmidt},
  \& {Schneider}}]{Dalcanton97}
{Dalcanton}, J.~J., {Spergel}, D.~N., {Gunn}, J.~E., {Schmidt}, M., \&
  {Schneider}, D.~P. 1997, \aj, 114, 635, \dodoi{10.1086/118499}

\bibitem[{{Danieli} {et~al.}(2019){Danieli}, {van Dokkum}, {Conroy}, {Abraham},
  \& {Romanowsky}}]{Danieli19}
{Danieli}, S., {van Dokkum}, P., {Conroy}, C., {Abraham}, R., \& {Romanowsky},
  A.~J. 2019, \apjl, 874, L12, \dodoi{10.3847/2041-8213/ab0e8c}

\bibitem[{{Danieli} {et~al.}(2017){Danieli}, {van Dokkum}, {Merritt},
  {Abraham}, {Zhang}, {Karachentsev}, \& {Makarova}}]{Danieli17}
{Danieli}, S., {van Dokkum}, P., {Merritt}, A., {et~al.} 2017, \apj, 837, 136,
  \dodoi{10.3847/1538-4357/aa615b}

\bibitem[{{Di Cintio} {et~al.}(2017){Di Cintio}, {Brook}, {Dutton},
  {Macci{\`o}}, {Obreja}, \& {Dekel}}]{Reff}
{Di Cintio}, A., {Brook}, C.~B., {Dutton}, A.~A., {et~al.} 2017, \mnras, 466,
  L1, \dodoi{10.1093/mnrasl/slw210}

\bibitem[{{Disney}(1976)}]{Disney76}
{Disney}, M.~J. 1976, \nat, 263, 573, \dodoi{10.1038/263573a0}

\bibitem[{{Doppel} {et~al.}(2021){Doppel}, {Sales}, {Navarro}, {Abadi}, {Peng},
  {Toloba}, \& {Ramos-Almendares}}]{Doppel21}
{Doppel}, J.~E., {Sales}, L.~V., {Navarro}, J.~F., {et~al.} 2021, \mnras, 502,
  1661, \dodoi{10.1093/mnras/staa3915}

\bibitem[{{Dubois} {et~al.}(2014){Dubois}, {Pichon}, {Welker}, {Le Borgne},
  {Devriendt}, {Laigle}, {Codis}, {Pogosyan}, {Arnouts}, {Benabed}, {Bertin},
  {Blaizot}, {Bouchet}, {Cardoso}, {Colombi}, {de Lapparent}, {Desjacques},
  {Gavazzi}, {Kassin}, {Kimm}, {McCracken}, {Milliard}, {Peirani}, {Prunet},
  {Rouberol}, {Silk}, {Slyz}, {Sousbie}, {Teyssier}, {Tresse}, {Treyer},
  {Vibert}, \& {Volonteri}}]{Dubois14}
{Dubois}, Y., {Pichon}, C., {Welker}, C., {et~al.} 2014, \mnras, 444, 1453,
  \dodoi{10.1093/mnras/stu1227}

\bibitem[{{Dubois} {et~al.}(2021){Dubois}, {Beckmann}, {Bournaud}, {Choi},
  {Devriendt}, {Jackson}, {Kaviraj}, {Kimm}, {Kraljic}, {Laigle}, {Martin},
  {Park}, {Peirani}, {Pichon}, {Volonteri}, \& {Yi}}]{Dubois21}
{Dubois}, Y., {Beckmann}, R., {Bournaud}, F., {et~al.} 2021, \aap, 651, A109,
  \dodoi{10.1051/0004-6361/202039429}

\bibitem[{{Dutta Chowdhury} {et~al.}(2019){Dutta Chowdhury}, {van den Bosch},
  \& {van Dokkum}}]{Chowdhury19}
{Dutta Chowdhury}, D., {van den Bosch}, F.~C., \& {van Dokkum}, P. 2019, arXiv
  e-prints, arXiv:1902.05959.
\newblock \doarXiv{1902.05959}

\bibitem[{{Dutton} {et~al.}(2020){Dutton}, {Buck}, {Macci{\`o}}, {Dixon},
  {Blank}, \& {Obreja}}]{Dutton2020}
{Dutton}, A.~A., {Buck}, T., {Macci{\`o}}, A.~V., {et~al.} 2020, \mnras, 499,
  2648, \dodoi{10.1093/mnras/staa3028}

\bibitem[{{Ferr{\'e}-Mateu} {et~al.}(2018){Ferr{\'e}-Mateu}, {Alabi}, {Forbes},
  {Romanowsky}, {Brodie}, {Pandya}, {Mart{\'\i}n-Navarro}, {Bellstedt},
  {Wasserman}, {Stone}, \& {Okabe}}]{Ferremateu18}
{Ferr{\'e}-Mateu}, A., {Alabi}, A., {Forbes}, D.~A., {et~al.} 2018, \mnras,
  479, 4891, \dodoi{10.1093/mnras/sty1597}

\bibitem[{{Forbes} {et~al.}(2020){Forbes}, {Alabi}, {Romanowsky}, {Brodie}, \&
  {Arimoto}}]{Forbes20}
{Forbes}, D.~A., {Alabi}, A., {Romanowsky}, A.~J., {Brodie}, J.~P., \&
  {Arimoto}, N. 2020, \mnras, 492, 4874, \dodoi{10.1093/mnras/staa180}

\bibitem[{{Geha} {et~al.}(2012){Geha}, {Blanton}, {Yan}, \& {Tinker}}]{Geha12}
{Geha}, M., {Blanton}, M.~R., {Yan}, R., \& {Tinker}, J.~L. 2012, \apj, 757,
  85, \dodoi{10.1088/0004-637X/757/1/85}

\bibitem[{{Governato} {et~al.}(2015){Governato}, {Weisz}, {Pontzen}, {Loebman},
  {Reed}, {Brooks}, {Behroozi}, {Christensen}, {Madau}, {Mayer}, {Shen},
  {Walker}, {Quinn}, {Keller}, \& {Wadsley}}]{Governato15}
{Governato}, F., {Weisz}, D., {Pontzen}, A., {et~al.} 2015, \mnras, 448, 792,
  \dodoi{10.1093/mnras/stu2720}

\bibitem[{{Greco} {et~al.}(2018){Greco}, {Greene}, {Strauss}, {Macarthur},
  {Flowers}, {Goulding}, {Huang}, {Kim}, {Komiyama}, {Leauthaud}, {Leisman},
  {Lupton}, {Sif{\'o}n}, \& {Wang}}]{Greco18}
{Greco}, J.~P., {Greene}, J.~E., {Strauss}, M.~A., {et~al.} 2018, \apj, 857,
  104, \dodoi{10.3847/1538-4357/aab842}

\bibitem[{{Haardt} \& {Madau}(2012)}]{Haardt12}
{Haardt}, F., \& {Madau}, P. 2012, \apj, 746, 125,
  \dodoi{10.1088/0004-637X/746/2/125}

\bibitem[{{Hetznecker} \& {Burkert}(2006)}]{Hetznecker06}
{Hetznecker}, H., \& {Burkert}, A. 2006, \mnras, 370, 1905,
  \dodoi{10.1111/j.1365-2966.2006.10616.x}

\bibitem[{{Hunter}(2007)}]{Hunter07}
{Hunter}, J.~D. 2007, Computing in Science and Engineering, 9, 90,
  \dodoi{10.1109/MCSE.2007.55}

\bibitem[{{Impey} {et~al.}(1988){Impey}, {Bothun}, \& {Malin}}]{Impey88}
{Impey}, C., {Bothun}, G., \& {Malin}, D. 1988, \apj, 330, 634,
  \dodoi{10.1086/166500}

\bibitem[{{Jackson} {et~al.}(2020){Jackson}, {Martin}, {Kaviraj}, {Rams{\o}y},
  {Devriendt}, {Sedgwick}, {Laigle}, {Choi}, {Beckmann}, {Volonteri}, {Dubois},
  {Pichon}, {Yi}, {Slyz}, {Kraljic}, {Kimm}, {Peirani}, \&
  {Baldry}}]{Jackson20}
{Jackson}, R.~A., {Martin}, G., {Kaviraj}, S., {et~al.} 2020, arXiv e-prints,
  arXiv:2007.06581.
\newblock \doarXiv{2007.06581}

\bibitem[{{Jester} {et~al.}(2005){Jester}, {Schneider}, {Richards}, {Green},
  {Schmidt}, {Hall}, {Strauss}, {Vand en Berk}, {Stoughton}, {Gunn},
  {Brinkmann}, {Kent}, {Smith}, {Tucker}, \& {Yanny}}]{Jester}
{Jester}, S., {Schneider}, D.~P., {Richards}, G.~T., {et~al.} 2005, \aj, 130,
  873, \dodoi{10.1086/432466}

\bibitem[{{Jiang} {et~al.}(2019){Jiang}, {Dekel}, {Freundlich}, {Romanowsky},
  {Dutton}, {Macci{\`o}}, \& {Di Cintio}}]{Jiang19}
{Jiang}, F., {Dekel}, A., {Freundlich}, J., {et~al.} 2019, \mnras, 487, 5272,
  \dodoi{10.1093/mnras/stz1499}

\bibitem[{{Kado-Fong} {et~al.}(2021){Kado-Fong}, {Petrescu}, {Mohammad},
  {Greco}, {Greene}, {Adams}, {Huang}, {Leisman}, {Munshi}, {Tanoglidis}, \&
  {Van Nest}}]{KadoFong21}
{Kado-Fong}, E., {Petrescu}, M., {Mohammad}, M., {et~al.} 2021, arXiv e-prints,
  arXiv:2106.05288.
\newblock \doarXiv{2106.05288}

\bibitem[{{Katz} \& {White}(1993)}]{ZoomReNorm}
{Katz}, N., \& {White}, S. D.~M. 1993, \apj, 412, 455, \dodoi{10.1086/172935}

\bibitem[{{Knollmann} \& {Knebe}(2009)}]{AHF}
{Knollmann}, S.~R., \& {Knebe}, A. 2009, \apjs, 182, 608,
  \dodoi{10.1088/0067-0049/182/2/608}

\bibitem[{{Koda} {et~al.}(2015){Koda}, {Yagi}, {Yamanoi}, \&
  {Komiyama}}]{Koda15}
{Koda}, J., {Yagi}, M., {Yamanoi}, H., \& {Komiyama}, Y. 2015, \apjl, 807, L2,
  \dodoi{10.1088/2041-8205/807/1/L2}

\bibitem[{{Kroupa}(2001)}]{Kroupa01}
{Kroupa}, P. 2001, \mnras, 322, 231, \dodoi{10.1046/j.1365-8711.2001.04022.x}

\bibitem[{{Leisman} {et~al.}(2017){Leisman}, {Haynes}, {Janowiecki},
  {Hallenbeck}, {J{\'o}zsa}, {Giovanelli}, {Adams}, {Bernal Neira}, {Cannon},
  {Janesh}, {Rhode}, \& {Salzer}}]{Leisman17}
{Leisman}, L., {Haynes}, M.~P., {Janowiecki}, S., {et~al.} 2017, \apj, 842,
  133, \dodoi{10.3847/1538-4357/aa7575}

\bibitem[{{Liao} {et~al.}(2019){Liao}, {Gao}, {Frenk}, {Grand}, {Guo},
  {G{\'o}mez}, {Marinacci}, {Pakmor}, {Shao}, \& {Springel}}]{Liao19}
{Liao}, S., {Gao}, L., {Frenk}, C.~S., {et~al.} 2019, \mnras, 490, 5182,
  \dodoi{10.1093/mnras/stz2969}

\bibitem[{{Ludlow} {et~al.}(2019){Ludlow}, {Schaye}, {Schaller}, \&
  {Richings}}]{Ludlow}
{Ludlow}, A.~D., {Schaye}, J., {Schaller}, M., \& {Richings}, J. 2019, \mnras,
  488, L123, \dodoi{10.1093/mnrasl/slz110}

\bibitem[{{Martin} {et~al.}(2019){Martin}, {Kaviraj}, {Laigle}, {Devriendt},
  {Jackson}, {Peirani}, {Dubois}, {Pichon}, \& {Slyz}}]{Martin19}
{Martin}, G., {Kaviraj}, S., {Laigle}, C., {et~al.} 2019, \mnras, 485, 796,
  \dodoi{10.1093/mnras/stz356}

\bibitem[{{Menon} {et~al.}(2015){Menon}, {Wesolowski}, {Zheng}, {Jetley},
  {Kale}, {Quinn}, \& {Governato}}]{Changa}
{Menon}, H., {Wesolowski}, L., {Zheng}, G., {et~al.} 2015, Computational
  Astrophysics and Cosmology, 2, 1, \dodoi{10.1186/s40668-015-0007-9}

\bibitem[{{Mihos} {et~al.}(2015){Mihos}, {Durrell}, {Ferrarese}, {Feldmeier},
  {C{\^o}t{\'e}}, {Peng}, {Harding}, {Liu}, {Gwyn}, \& {Cuillandre}}]{Mihos15}
{Mihos}, J.~C., {Durrell}, P.~R., {Ferrarese}, L., {et~al.} 2015, \apj, 809,
  L21, \dodoi{10.1088/2041-8205/809/2/L21}

\bibitem[{{Moster} {et~al.}(2013){Moster}, {Naab}, \& {White}}]{SM-HM}
{Moster}, B.~P., {Naab}, T., \& {White}, S. D.~M. 2013, \mnras, 428, 3121,
  \dodoi{10.1093/mnras/sts261}

\bibitem[{{Munshi} {et~al.}(2021){Munshi}, {Brooks}, {Applebaum},
  {Christensen}, {Sligh}, \& {Quinn}}]{Munshi21}
{Munshi}, F., {Brooks}, A., {Applebaum}, E., {et~al.} 2021, arXiv e-prints,
  arXiv:2101.05822.
\newblock \doarXiv{2101.05822}

\bibitem[{{Munshi} {et~al.}(2013){Munshi}, {Governato}, {Brooks},
  {Christensen}, {Shen}, {Loebman}, {Moster}, {Quinn}, \& {Wadsley}}]{Munshi13}
{Munshi}, F., {Governato}, F., {Brooks}, A.~M., {et~al.} 2013, \apj, 766, 56,
  \dodoi{10.1088/0004-637X/766/1/56}

\bibitem[{{Ogiya}(2018)}]{Ogiya18}
{Ogiya}, G. 2018, \mnras, 480, L106, \dodoi{10.1093/mnrasl/sly138}

\bibitem[{{Papastergis} {et~al.}(2017){Papastergis}, {Adams}, \&
  {Romanowsky}}]{Papastergis2017}
{Papastergis}, E., {Adams}, E.~A.~K., \& {Romanowsky}, A.~J. 2017, \aap, 601,
  L10, \dodoi{10.1051/0004-6361/201730795}

\bibitem[{{Planck Collaboration} {et~al.}(2014){Planck Collaboration}, {Ade},
  {Aghanim}, {Armitage-Caplan}, {Arnaud}, {Ashdown}, {Atrio-Barand ela},
  {Aumont}, {Baccigalupi}, {Banday}, {Barreiro}, {Bartlett}, {Battaner},
  {Benabed}, {Beno{\^\i}t}, {Benoit-L{\'e}vy}, {Bernard}, {Bersanelli},
  {Bielewicz}, {Bobin}, {Bock}, {Bonaldi}, {Bond}, {Borrill}, {Bouchet},
  {Bridges}, {Bucher}, {Burigana}, {Butler}, {Calabrese}, {Cappellini},
  {Cardoso}, {Catalano}, {Challinor}, {Chamballu}, {Chary}, {Chen}, {Chiang},
  {Chiang}, {Christensen}, {Church}, {Clements}, {Colombi}, {Colombo},
  {Couchot}, {Coulais}, {Crill}, {Curto}, {Cuttaia}, {Danese}, {Davies},
  {Davis}, {de Bernardis}, {de Rosa}, {de Zotti}, {Delabrouille}, {Delouis},
  {D{\'e}sert}, {Dickinson}, {Diego}, {Dolag}, {Dole}, {Donzelli}, {Dor{\'e}},
  {Douspis}, {Dunkley}, {Dupac}, {Efstathiou}, {Elsner}, {En{\ss}lin},
  {Eriksen}, {Finelli}, {Forni}, {Frailis}, {Fraisse}, {Franceschi}, {Gaier},
  {Galeotta}, {Galli}, {Ganga}, {Giard}, {Giardino}, {Giraud-H{\'e}raud},
  {Gjerl{\o}w}, {Gonz{\'a}lez-Nuevo}, {G{\'o}rski}, {Gratton}, {Gregorio},
  {Gruppuso}, {Gudmundsson}, {Haissinski}, {Hamann}, {Hansen}, {Hanson},
  {Harrison}, {Henrot-Versill{\'e}}, {Hern{\'a}ndez-Monteagudo}, {Herranz},
  {Hildebrand t}, {Hivon}, {Hobson}, {Holmes}, {Hornstrup}, {Hou}, {Hovest},
  {Huffenberger}, {Jaffe}, {Jaffe}, {Jewell}, {Jones}, {Juvela},
  {Keih{\"a}nen}, {Keskitalo}, {Kisner}, {Kneissl}, {Knoche}, {Knox}, {Kunz},
  {Kurki-Suonio}, {Lagache}, {L{\"a}hteenm{\"a}ki}, {Lamarre}, {Lasenby},
  {Lattanzi}, {Laureijs}, {Lawrence}, {Leach}, {Leahy}, {Leonardi},
  {Le{\'o}n-Tavares}, {Lesgourgues}, {Lewis}, {Liguori}, {Lilje},
  {Linden-V{\o}rnle}, {L{\'o}pez-Caniego}, {Lubin}, {Mac{\'\i}as-P{\'e}rez},
  {Maffei}, {Maino}, {Mand olesi}, {Maris}, {Marshall}, {Martin},
  {Mart{\'\i}nez-Gonz{\'a}lez}, {Masi}, {Massardi}, {Matarrese}, {Matthai},
  {Mazzotta}, {Meinhold}, {Melchiorri}, {Melin}, {Mendes}, {Menegoni},
  {Mennella}, {Migliaccio}, {Millea}, {Mitra}, {Miville-Desch{\^e}nes},
  {Moneti}, {Montier}, {Morgante}, {Mortlock}, {Moss}, {Munshi}, {Murphy},
  {Naselsky}, {Nati}, {Natoli}, {Netterfield}, {N{\o}rgaard-Nielsen},
  {Noviello}, {Novikov}, {Novikov}, {O'Dwyer}, {Osborne}, {Oxborrow}, {Paci},
  {Pagano}, {Pajot}, {Paladini}, {Paoletti}, {Partridge}, {Pasian},
  {Patanchon}, {Pearson}, {Pearson}, {Peiris}, {Perdereau}, {Perotto},
  {Perrotta}, {Pettorino}, {Piacentini}, {Piat}, {Pierpaoli}, {Pietrobon},
  {Plaszczynski}, {Platania}, {Pointecouteau}, {Polenta}, {Ponthieu}, {Popa},
  {Poutanen}, {Pratt}, {Pr{\'e}zeau}, {Prunet}, {Puget}, {Rachen}, {Reach},
  {Rebolo}, {Reinecke}, {Remazeilles}, {Renault}, {Ricciardi}, {Riller},
  {Ristorcelli}, {Rocha}, {Rosset}, {Roudier}, {Rowan-Robinson},
  {Rubi{\~n}o-Mart{\'\i}n}, {Rusholme}, {Sandri}, {Santos}, {Savelainen},
  {Savini}, {Scott}, {Seiffert}, {Shellard}, {Spencer}, {Starck}, {Stolyarov},
  {Stompor}, {Sudiwala}, {Sunyaev}, {Sureau}, {Sutton}, {Suur-Uski}, {Sygnet},
  {Tauber}, {Tavagnacco}, {Terenzi}, {Toffolatti}, {Tomasi}, {Tristram},
  {Tucci}, {Tuovinen}, {T{\"u}rler}, {Umana}, {Valenziano}, {Valiviita}, {Van
  Tent}, {Vielva}, {Villa}, {Vittorio}, {Wade}, {Wandelt}, {Wehus}, {White},
  {White}, {Wilkinson}, {Yvon}, {Zacchei}, \& {Zonca}}]{Planck}
{Planck Collaboration}, {Ade}, P.~A.~R., {Aghanim}, N., {et~al.} 2014, \aap,
  571, A16, \dodoi{10.1051/0004-6361/201321591}

\bibitem[{{Pontzen} {et~al.}(2013){Pontzen}, {Ro{\v s}kar}, {Stinson}, \&
  {Woods}}]{pynbody}
{Pontzen}, A., {Ro{\v s}kar}, R., {Stinson}, G., \& {Woods}, R. 2013, {pynbody:
  N-Body/SPH analysis for python}, Astrophysics Source Code Library, record
  ascl 1305.002.
\newblock \doeprint{1305.002}

\bibitem[{{Pontzen} \& {Tremmel}(2018)}]{Tangos}
{Pontzen}, A., \& {Tremmel}, M. 2018, \apjs, 237, 23,
  \dodoi{10.3847/1538-4365/aac832}

\bibitem[{{Pontzen} {et~al.}(2008){Pontzen}, {Governato}, {Pettini}, {Booth},
  {Stinson}, {Wadsley}, {Brooks}, {Quinn}, \& {Haehnelt}}]{Pontzen08}
{Pontzen}, A., {Governato}, F., {Pettini}, M., {et~al.} 2008, \mnras, 390,
  1349, \dodoi{10.1111/j.1365-2966.2008.13782.x}

\bibitem[{{Rong} {et~al.}(2017){Rong}, {Guo}, {Gao}, {Liao}, {Xie}, {Puzia},
  {Sun}, \& {Pan}}]{Rong19}
{Rong}, Y., {Guo}, Q., {Gao}, L., {et~al.} 2017, \mnras, 470, 4231,
  \dodoi{10.1093/mnras/stx1440}

\bibitem[{{Rong} {et~al.}(2020){Rong}, {Dong}, {Puzia}, {Galaz},
  {S{\'a}nchez-Janssen}, {Cao}, {van der Burg}, {Sif{\'o}n}, {Mancera
  Pi{\~n}a}, {Marcelo}, {D'Ago}, {Zhang}, {Johnston}, \&
  {Eigenthaler}}]{Rong20}
{Rong}, Y., {Dong}, X.-Y., {Puzia}, T.~H., {et~al.} 2020, \apj, 899, 78,
  \dodoi{10.3847/1538-4357/aba74a}

\bibitem[{{Saifollahi} {et~al.}(2020){Saifollahi}, {Trujillo}, {Beasley},
  {Peletier}, \& {Knapen}}]{Saifollahi20}
{Saifollahi}, T., {Trujillo}, I., {Beasley}, M.~A., {Peletier}, R.~F., \&
  {Knapen}, J.~H. 2020, \mnras, \dodoi{10.1093/mnras/staa3016}

\bibitem[{{Saitoh} \& {Makino}(2009)}]{Saitoh09}
{Saitoh}, T.~R., \& {Makino}, J. 2009, \apjl, 697, L99,
  \dodoi{10.1088/0004-637X/697/2/L99}

\bibitem[{{Sandage} \& {Binggeli}(1984)}]{Sandage84}
{Sandage}, A., \& {Binggeli}, B. 1984, \aj, 89, 919, \dodoi{10.1086/113588}

\bibitem[{{Schramm} \& {Silverman}(2013)}]{SM-SMBH}
{Schramm}, M., \& {Silverman}, J.~D. 2013, \apj, 767, 13,
  \dodoi{10.1088/0004-637X/767/1/13}

\bibitem[{{S{\'e}rsic}(1963)}]{Sersic}
{S{\'e}rsic}, J.~L. 1963, Boletin de la Asociacion Argentina de Astronomia La
  Plata Argentina, 6, 41

\bibitem[{{Shen} {et~al.}(2010){Shen}, {Wadsley}, \& {Stinson}}]{Shen10}
{Shen}, S., {Wadsley}, J., \& {Stinson}, G. 2010, \mnras, 407, 1581,
  \dodoi{10.1111/j.1365-2966.2010.17047.x}

\bibitem[{{Stinson} {et~al.}(2006){Stinson}, {Seth}, {Katz}, {Wadsley},
  {Governato}, \& {Quinn}}]{Stinson06}
{Stinson}, G., {Seth}, A., {Katz}, N., {et~al.} 2006, \mnras, 373, 1074,
  \dodoi{10.1111/j.1365-2966.2006.11097.x}

\bibitem[{{Tanoglidis} {et~al.}(2021){Tanoglidis}, {Drlica-Wagner}, {Wei},
  {Li}, {S{\'a}nchez}, {Zhang}, {Peter}, {Feldmeier-Krause}, {Prat}, {Casey},
  {Palmese}, {S{\'a}nchez}, {DeRose}, {Conselice}, {Gagnon}, {Abbott},
  {Aguena}, {Allam}, {Avila}, {Bechtol}, {Bertin}, {Bhargava}, {Brooks},
  {Burke}, {Rosell}, {Kind}, {Carretero}, {Chang}, {Costanzi}, {da Costa}, {De
  Vicente}, {Desai}, {Diehl}, {Doel}, {Eifler}, {Everett}, {Evrard},
  {Flaugher}, {Frieman}, {Garc{\'\i}a-Bellido}, {Gerdes}, {Gruendl},
  {Gschwend}, {Gutierrez}, {Hartley}, {Hollowood}, {Huterer}, {James},
  {Krause}, {Kuehn}, {Kuropatkin}, {Maia}, {March}, {Marshall}, {Menanteau},
  {Miquel}, {Ogando}, {Paz-Chinch{\'o}n}, {Romer}, {Roodman}, {Sanchez},
  {Scarpine}, {Serrano}, {Sevilla-Noarbe}, {Smith}, {Suchyta}, {Tarle},
  {Thomas}, {Tucker}, {Walker}, \& {DES Collaboration}}]{Tanoglidis21}
{Tanoglidis}, D., {Drlica-Wagner}, A., {Wei}, K., {et~al.} 2021, \apjs, 252,
  18, \dodoi{10.3847/1538-4365/abca89}

\bibitem[{{Tremmel} {et~al.}(2018{\natexlab{a}}){Tremmel}, {Governato},
  {Volonteri}, {Pontzen}, \& {Quinn}}]{Tremmel18b}
{Tremmel}, M., {Governato}, F., {Volonteri}, M., {Pontzen}, A., \& {Quinn},
  T.~R. 2018{\natexlab{a}}, \apjl, 857, L22, \dodoi{10.3847/2041-8213/aabc0a}

\bibitem[{{Tremmel} {et~al.}(2015){Tremmel}, {Governato}, {Volonteri}, \&
  {Quinn}}]{TremmelSMBH}
{Tremmel}, M., {Governato}, F., {Volonteri}, M., \& {Quinn}, T.~R. 2015,
  \mnras, 451, 1868, \dodoi{10.1093/mnras/stv1060}

\bibitem[{{Tremmel} {et~al.}(2018{\natexlab{b}}){Tremmel}, {Governato},
  {Volonteri}, {Quinn}, \& {Pontzen}}]{Tremmel18a}
{Tremmel}, M., {Governato}, F., {Volonteri}, M., {Quinn}, T.~R., \& {Pontzen},
  A. 2018{\natexlab{b}}, \mnras, 475, 4967, \dodoi{10.1093/mnras/sty139}

\bibitem[{{Tremmel} {et~al.}(2017){Tremmel}, {Karcher}, {Governato},
  {Volonteri}, {Quinn}, {Pontzen}, {Anderson}, \& {Bellovary}}]{Rom25}
{Tremmel}, M., {Karcher}, M., {Governato}, F., {et~al.} 2017, \mnras, 470,
  1121, \dodoi{10.1093/mnras/stx1160}

\bibitem[{{Tremmel} {et~al.}(2020){Tremmel}, {Wright}, {Brooks}, {Munshi},
  {Nagai}, \& {Quinn}}]{Tremmel20}
{Tremmel}, M., {Wright}, A.~C., {Brooks}, A.~M., {et~al.} 2020, \mnras, 497,
  2786, \dodoi{10.1093/mnras/staa2015}

\bibitem[{{Tremmel} {et~al.}(2019){Tremmel}, {Quinn}, {Ricarte}, {Babul},
  {Chadayammuri}, {Natarajan}, {Nagai}, {Pontzen}, \& {Volonteri}}]{RomC}
{Tremmel}, M., {Quinn}, T.~R., {Ricarte}, A., {et~al.} 2019, \mnras, 483, 3336,
  \dodoi{10.1093/mnras/sty3336}

\bibitem[{{Trujillo} {et~al.}(2020){Trujillo}, {Chamba}, \&
  {Knapen}}]{Trujillo20}
{Trujillo}, I., {Chamba}, N., \& {Knapen}, J.~H. 2020, \mnras, 493, 87,
  \dodoi{10.1093/mnras/staa236}

\bibitem[{{Trujillo} \& {Fliri}(2016)}]{Trujillo16}
{Trujillo}, I., \& {Fliri}, J. 2016, \apj, 823, 123,
  \dodoi{10.3847/0004-637X/823/2/123}

\bibitem[{{van der Burg} {et~al.}(2016){van der Burg}, {Muzzin}, \&
  {Hoekstra}}]{SBavg}
{van der Burg}, R. F.~J., {Muzzin}, A., \& {Hoekstra}, H. 2016, \aap, 590, A20,
  \dodoi{10.1051/0004-6361/201628222}

\bibitem[{{van der Burg} {et~al.}(2017){van der Burg}, {Hoekstra}, {Muzzin},
  {Sif{\'o}n}, {Viola}, {Bremer}, {Brough}, {Driver}, {Erben}, {Heymans},
  {Hildebrandt}, {Holwerda}, {Klaes}, {Kuijken}, {McGee}, {Nakajima},
  {Napolitano}, {Norberg}, {Taylor}, \& {Valentijn}}]{vanderBurg17}
{van der Burg}, R. F.~J., {Hoekstra}, H., {Muzzin}, A., {et~al.} 2017, \aap,
  607, A79, \dodoi{10.1051/0004-6361/201731335}

\bibitem[{{van der Walt} {et~al.}(2011){van der Walt}, {Colbert}, \&
  {Varoquaux}}]{vanderWalt11}
{van der Walt}, S., {Colbert}, S.~C., \& {Varoquaux}, G. 2011, Computing in
  Science and Engineering, 13, 22, \dodoi{10.1109/MCSE.2011.37}

\bibitem[{{van Dokkum} {et~al.}(2019){van Dokkum}, {Danieli}, {Abraham},
  {Conroy}, \& {Romanowsky}}]{vanDokkum19}
{van Dokkum}, P., {Danieli}, S., {Abraham}, R., {Conroy}, C., \& {Romanowsky},
  A.~J. 2019, \apjl, 874, L5, \dodoi{10.3847/2041-8213/ab0d92}

\bibitem[{{van Dokkum} {et~al.}(2017){van Dokkum}, {Abraham}, {Romanowsky},
  {Brodie}, {Conroy}, {Danieli}, {Lokhorst}, {Merritt}, {Mowla}, \&
  {Zhang}}]{vanDokkum17}
{van Dokkum}, P., {Abraham}, R., {Romanowsky}, A.~J., {et~al.} 2017, \apjl,
  844, L11, \dodoi{10.3847/2041-8213/aa7ca2}

\bibitem[{{van Dokkum} {et~al.}(2018){van Dokkum}, {Danieli}, {Cohen},
  {Merritt}, {Romanowsky}, {Abraham}, {Brodie}, {Conroy}, {Lokhorst}, {Mowla},
  {O'Sullivan}, \& {Zhang}}]{vanDokkum18}
{van Dokkum}, P., {Danieli}, S., {Cohen}, Y., {et~al.} 2018, \nat, 555, 629,
  \dodoi{10.1038/nature25767}

\bibitem[{{van Dokkum} {et~al.}(2015){van Dokkum}, {Abraham}, {Merritt},
  {Zhang}, {Geha}, \& {Conroy}}]{Gband}
{van Dokkum}, P.~G., {Abraham}, R., {Merritt}, A., {et~al.} 2015, \apjl, 798,
  L45, \dodoi{10.1088/2041-8205/798/2/L45}

\bibitem[{{Venhola} {et~al.}(2017){Venhola}, {Peletier}, {Laurikainen}, {Salo},
  {Lisker}, {Iodice}, {Capaccioli}, {Verdois Kleijn}, {Valentijn}, \&
  {Mieske}}]{Venhola17}
{Venhola}, A., {Peletier}, R., {Laurikainen}, E., {et~al.} 2017, \aap, 608,
  A142, \dodoi{10.1051/0004-6361/201730696}

\bibitem[{{Verner} \& {Ferland}(1996)}]{Verner96}
{Verner}, D.~A., \& {Ferland}, G.~J. 1996, \apjs, 103, 467,
  \dodoi{10.1086/192284}

\bibitem[{{Wadsley} {et~al.}(2017){Wadsley}, {Keller}, \& {Quinn}}]{Gasoline2}
{Wadsley}, J.~W., {Keller}, B.~W., \& {Quinn}, T.~R. 2017, \mnras, 471, 2357,
  \dodoi{10.1093/mnras/stx1643}

\bibitem[{{Wadsley} {et~al.}(2004){Wadsley}, {Stadel}, \& {Quinn}}]{Gasoline}
{Wadsley}, J.~W., {Stadel}, J., \& {Quinn}, T. 2004, \na, 9, 137,
  \dodoi{10.1016/j.newast.2003.08.004}

\bibitem[{{Wadsley} {et~al.}(2008){Wadsley}, {Veeravalli}, \&
  {Couchman}}]{Wadsley08}
{Wadsley}, J.~W., {Veeravalli}, G., \& {Couchman}, H.~M.~P. 2008, \mnras, 387,
  427, \dodoi{10.1111/j.1365-2966.2008.13260.x}

\bibitem[{{Wright} {et~al.}(2021){Wright}, {Tremmel}, {Brooks}, {Munshi},
  {Nagai}, {Sharma}, \& {Quinn}}]{Wright21}
{Wright}, A.~C., {Tremmel}, M., {Brooks}, A.~M., {et~al.} 2021, \mnras, 502,
  5370, \dodoi{10.1093/mnras/stab081}

\bibitem[{{Yozin} \& {Bekki}(2015)}]{Yozin15}
{Yozin}, C., \& {Bekki}, K. 2015, \mnras, 452, 937,
  \dodoi{10.1093/mnras/stv1073}

\end{thebibliography}
\bibliographystyle{aasjournal}

\end{document}